\documentclass[11pt,a4paper]{article}
\usepackage{jcappub}
\usepackage[utf8]{inputenc}
\usepackage{babel}
\usepackage[dvipsnames]{xcolor}
\usepackage{array}
\usepackage{multirow}
\usepackage{amsmath,amssymb}
\usepackage{dsfont}
\usepackage{cancel}
\usepackage{soul}
\usepackage{stackrel}
\usepackage{braket}
\usepackage{physics}
\usepackage{fontawesome}
\usepackage{graphicx}
\usepackage{hhline}
\usepackage{tabularx}

\usepackage{colortbl} %Improved table environments

\usepackage{breakcites}
\usepackage[sort&compress]{natbib}
\usepackage[tmargin=2cm,bmargin=2.5cm,lmargin=2.8cm,rmargin=2.8cm]{geometry}
\hypersetup{unicode=true,pdfusetitle, bookmarks=true,bookmarksnumbered=false,bookmarksopen=true, breaklinks=false,backref=false,colorlinks=false}
\newcommand{\nede}{_{_\text{NEDE}}}

\begin{document}
	\title{A grounded perspective on New Early Dark Energy\\
	using ACT, SPT, and BICEP/Keck}
	\author[a]{Juan S. Cruz}
%	\email{jcr@sdu.dk}
	\emailAdd{jcr@sdu.dk}
	\affiliation[a]{CP3-Origins, Center for Cosmology and Particle Physics Phenomenology, University of Southern Denmark, Campusvej 55, 5230 Odense M, Denmark}

	\author[b]{Florian Niedermann}
%	\email{florian.niedermann@su.se}
	\emailAdd{florian.niedermann@su.se}
	\affiliation[b]{Nordita, KTH Royal Institute of Technology and Stockholm University Hannes Alfv\'ens v\"ag 12, SE-106 91 Stockholm, Sweden}

	\author[a]{Martin S. Sloth}
%	\email{sloth@cp3.sdu.dk}
	\emailAdd{sloth@cp3.sdu.dk}
%	\affiliation{CP3-Origins, Center for Cosmology and Particle Physics Phenomenology, University of Southern Denmark, Campusvej 55, 5230 Odense M, Denmark}


	\abstract{We examine further the ability of the New Early Dark Energy model (NEDE) to resolve the current tension between the Cosmic Microwave Background (CMB) and local measurements of $H_0$ and the consequences for inflation. We perform new Bayesian analyses, including the current datasets from the ground-based CMB telescopes Atacama Cosmology Telescope (ACT), the South Pole Telescope (SPT), and the BICEP/Keck telescopes, employing an updated likelihood for the local measurements coming from the S$H_0$ES collaboration. Using the S$H_0$ES prior on $H_0$, the combined analysis with Baryonic Acoustic Oscillations (BAO), Pantheon, Planck and ACT improves the best-fit by $\Delta\chi^2 =  -15.9$ with respect to $\Lambda$CDM, favors a non-zero fractional contribution of NEDE, $f\nede > 0$, by $4.8\sigma$, and gives a best-fit value for the Hubble constant of $H_0 = 72.09$ km/s/Mpc (mean $71.49\pm 0.82$ with $68\%$ C.L.). A similar analysis using SPT instead of ACT yields consistent results with a $\Delta \chi^2 = - 23.1$ over $\Lambda$CDM, a preference for non-zero $f\nede$ of $4.7\sigma$ and a best-fit value of $H_0=71.77$ km/s/Mpc (mean $71.43\pm 0.85$ with $68\%$ C.L.). We also provide the constraints on the inflation parameters $r$ and $n_s$ coming from NEDE, including the BICEP/Keck 2018 data, and show that the allowed upper value on the tensor-scalar ratio is consistent with the $\Lambda$CDM bound, but, as also originally found, with a more blue scalar spectrum implying that the simplest curvaton model is now favored over the Starobinsky inflation model.}

	\maketitle

\vspace*{.7cm}
\begin{flushright}
\emph{``Keep your eyes on the stars, but your feet on the ground."}

---Theodore Roosevelt
\end{flushright}

\section{Introduction}

The current standard model of cosmology is the $\Lambda$CDM model. It portrays the contents of the Universe as perfect fluids, with the different components observing particular equations of state. These are specifically: a cosmological constant, cold dark matter, baryonic matter, and radiation. The model is the result of combining general relativity and the cosmological principle, i.e.\ the assumption that the Universe is homogeneous and isotropic, and specifying the energy-momentum tensor as a sum of the aforementioned components. The first two give out the FLRW type metric, which features a scale factor and allows the Hubble parameter to be defined. However, only after all these ingredients are included does one has a complete cosmological model.

$\Lambda$CDM is simple enough as to be described by six free parameters (assuming no spatial curvature): $\{\omega_b,\omega_c, \theta_*,\tau_{\rm reio}, A_s, n_s \}$, where the last two are related to the inflationary era. Since its introduction, it has been extremely successful at explaining the Universe we observe, but a new generation of measurements is putting it to the test~\cite{Nadathur:2020kvq,Abdalla:2022yfr}. Different current experiments show growing tensions in some of the parameters derived from $\Lambda$CDM. The largest tension so far is that of the Hubble parameter at present time, $H_0$ \cite{Knox:2019rjx,DiValentino:2020zio}.

The tension arises from the disagreement between values inferred from the CMB at high redshifts, most notoriously $H_0= 67.27\pm 0.60$ km/s/Mpc from the Planck collaboration \cite{Planck:2018vyg}, and those from local measurements at lower redshifts, which are mostly obtained, but not only, by constructing a distance ladder using standard candles. The S$H_0$ES collaboration \cite{Riess:2021jrx} stands out among them, reporting a value of $73.04\pm 1.04$ km/s/Mpc. There are many more measurements on both categories still in line with the trend of an early- versus late-time discrepancy, however, with larger errors. It is worth mentioning that at low values of redshift, the dependence of these measurements on the $\Lambda$CDM model is only mild, while the CMB measurement, although extremely precise, relies heavily on the model to evolve the fitted parameters to its present values. These two measurements claim high enough precision for their mean values to present us with the so-called \emph{Hubble tension} at around $5\sigma$ or more, depending on the datasets combined. Local under densities and other systematics have already been considered, and so far, it is not possible to explain the tension as being a consequence of only systematic issues~\cite{Wojtak:2013gda, Odderskov:2017ivg, Wu:2017fpr, Davis:2019wet}

Assuming the reported tension is not simply due to unknown systematic errors, a modification of the underlying model and assumptions must be considered. Several models have been put out in an attempt to explain the tension (see \cite{DiValentino:2021izs} for a review), but few models truly reduce the tension over $\Lambda$CDM when all the data, including CMB, BAO, and SN data, are included. The most promising models have been compared with each other under reasonable criteria and ranked according to how well they reduce the Hubble tension in the cosmological data \cite{Schoneberg:2021qvd}. Future data will be able to better constrain and discriminate between the few top phenomenologically favored models. Thus, further improvement of the models and testing them against more data is currently an important task.

The Hubble tension is closely related to the sound horizon \cite{Bernal:2016gxb, Aylor:2018drw, Knox:2019rjx,Arendse:2019hev}. As with $H_0$, the size of the sound horizon at recombination, $r_s^{\rm rec}$, or alternatively at the baryon drag epoch, $r_s^{\rm d}$\footnote{They appear to differ by a few percent in a relatively model-independent way.}, is different when deduced from CMB measurements and when it is obtained from the classical distance ladder (CDL) together with BAO. The latter, when combined with Supernovae (SNe) data, fixes the product $H_0 r_s$ without assuming the $\Lambda$CDM model. This seems to point at solutions to the tension that change the cosmic evolution just prior to recombination.

Along with the Hubble tension, an additional tension in the cosmological parameter $\sigma_8$ has been observed. As explained in further detail in the reviews and papers \cite{DiValentino:2020vvd, Nunes:2021ipq, Asadi:2022njl, Abdalla:2022yfr}, there is a tension between $2\sigma$ and $3\sigma$ among low and high redshift measurements for the strength of matter clustering given by $S8\equiv \sigma_8\sqrt{\Omega_m/0.3}$. The parameter $\sigma_8$ measures the amplitude of the power spectrum at the scale of $8h^{-1}$ Mpc and is known to lie roughly between 0.7 and 1.0. Early time measurements seem to prefer higher values of $\sigma_8$~\cite{Joudaki:2016kym,Hildebrandt:2016iqg,Hildebrandt:2018yau,KiDS:2020suj} while local measurements favor smaller values for it~\cite{Planck:2018vyg}. A generic feature of [New] Early Dark Energy ([N]EDE) type models is the slight increase in the baryonic and cold dark matter which has been seen to raise the value of $\sigma_8$ with a small statistically insignificant amount over $\Lambda$CDM   \cite{Poulin:2018dzj,DAmico:2020ods, Niedermann:2020dwg,Niedermann:2020qbw,Murgia:2020ryi,Smith:2020rxx,Simon:2022adh}.

In this article we concentrate on one of the models proposed to address the Hubble tension by reducing the size of the sound horizon at recombination/last-scattering, the New Early Dark Energy (NEDE) model \cite{Niedermann:2019olb, Niedermann:2020dwg}. It has been successful in reducing the tension when compared against data from Planck, BAO, SN Pantheon, and S$H_0$ES while increasing slightly the $\sigma_8$ tension in its first simple implementation, also later referred to as Cold NEDE. A newer proposal\footnote{For other interesting developments of NEDE, see also \cite{Allali:2021azp, Freese:2021rjq}.}, the Hot NEDE model \cite{Niedermann:2021vgd,Niedermann:2021ijp} may lead to different conclusions and holds in it the potential to reduce the $\sigma_8$ tension as well, however, we do not consider Hot NEDE in the present study.  In~\cite{Niedermann:2020dwg,Schoneberg:2021qvd}, Cold NEDE has also been compared to its predecessor EDE \cite{Karwal:2016vyq,Poulin:2018dzj,Poulin:2018cxd,Smith:2019ihp}, and it has been previously tested against CMB ground base data from the Atacama Cosmology Telescope~\cite{Poulin:2021bjr}. Similarly, data from SPT and ACT has already been used to constrain the EDE model~\cite{LaPosta:2021pgm, Hill:2021yec, Poulin:2021bjr, Smith:2022hwi, Jiang:2022uyg}. The objective of this paper is to update the statistics of the Cold NEDE model by including more datasets in the analysis. Although our main focus is to make an update in relation to the more relevant Hubble tension, we also report on the favored $\sigma_8$ values. Specifically, we are performing the corresponding Bayesian analysis with data coming from ACT, SPT, and BICEP together with our baseline datasets consisting of BAO, SNe, and Planck data. Our approach is slightly different from previous studies~\cite{Poulin:2021bjr} in which the multi-modality of NEDE is examined by separating the mass range. We instead perform our analysis under the assumption of a fixed equation of state and arrive at similar conclusions.

In the original work, it was already shown that NEDE prefers a much bluer spectrum of the initial adiabatic primordial curvature perturbation from inflation, which has important consequences for our understanding of inflation \cite{Niedermann:2019olb, Niedermann:2020dwg}. However, here, using BICEP/keck 2018 data, we provide for the first time a full $n_s-r$ constraints plot for NEDE (although a parametric estimate of it was given earlier in \cite{DAmico:2021fhz}). It looks qualitatively similar to the findings made within old EDE \cite{Karwal:2016vyq, Poulin:2018dzj,Smith:2019ihp} and shows that a simple curvaton model \cite{Enqvist:2001zp, Lyth:2001nq, Moroi:2001ct} is now favored over the Starobinsky inflation model~\cite{Starobinsky:1980te}.

The present paper is organized as follows. In section \ref{sec:theModel} we summarize the model and establish the terminology and notation relevant for the rest of the document. In the next one, section \ref{sec:mcmcAnalysis}, we describe the details of the Monte Carlo Markov Chain (MCMC) analysis, and discuss our results. It is subdivided into three sections, one for each new dataset: ACT, SPT, and BICEP. In the last section, we gather the most relevant observations and comment on the prospects of the NEDE model.

\section{The NEDE model}
\label{sec:theModel}

We start by introducing a phenomenological NEDE model as a general formalism for describing a phase transition in the CMB that can capture the physics of different field theoretic implementations such as cold, hot, and hybrid NEDE. Its distinctive feature is a \textit{trigger mechanism} that induces a sudden decay of an early dark energy component $\rho\nede$ at (cosmological) time $t=t_*(\mathbf{x})$ (which in general is a function of the spatial coordinate $\mathbf{x}$). The NEDE fluid at the background level is then described in terms of a time-dependent equation of state parameter
\begin{equation}
	w\nede(t) = \begin{cases} -1 &\quad \text{for}\quad t<\bar{t}_* \\ w\nede(t) &\quad\text{for}\quad t\geq \bar{t}_*\end{cases},
	\label{eq:wNEDEevol}
\end{equation}
where $\bar{t}_* $ arises from decomposing  $t_*(\mathbf{x}) = \bar{t}_* + \delta t_*(\mathbf{x})$ into a background and fluctuation part.

Before the transition $\rho\nede$ behaves like a cosmological constant, i.e. $\rho\nede (t< \bar{t}_*) = \bar{\rho}\nede^* = \mathrm{const}$.  In particular, it leads to an energy injection in the cosmic fluid that lowers the sound horizon~$r_s$. Since the CMB measures the angular scale of $r_s$ with high precision, the reduction in $r_s$ needs to be compensated by shortening the distance to the last scattering surface, which is achieved by raising $H_0$. Provided the fraction of NEDE, $f\nede = \rho\nede^* / \rho_\mathrm{tot}^*$, is of order of $10 \%$ and the transition happens around matter-radiation equality, the increase in $H_0$ is large enough to reconcile the CMB-inferred value of $H_0$ with the locally measured one.

After the transition, $w\nede(t)$ is allowed to have a complicated time-dependence within the bounds $1/3 \leq w\nede(t \geq \bar{t}_*) \leq 1 $, corresponding to a decay  of $\rho\nede$ at least as fast as radiation as expected in the bubble coalescence phase after the transition \cite{Niedermann:2019olb,Niedermann:2020dwg} (see also \cite{Gouttenoire:2021jhk}). This quick decay of NEDE is crucial to preserving the fit to the CMB power spectrum and having a viable late-time phenomenology. To be specific, after decomposing $\rho\nede = \bar{\rho}\nede(t) + \delta\rho\nede(t,\mathbf{x})$, we have
\begin{align}\label{eq:rho_nede}
\bar{\rho}\nede(t) = \bar{\rho}\nede^* \exp{- 3 \int_{\bar{t}_*}^t  \mathrm{d} \tilde t  H \left[1 + w\nede(\tilde t)\right]}\,,
\end{align}
where $H=\dot a / a$ is the Hubble parameter and we solved the energy conservation equation,  $\dot{\bar{\rho}}\nede + 3 H (1 + w\nede) \bar{\rho}\nede = 0$, subject to the boundary condition $\bar{\rho}\nede(\bar{t}_*) = \bar{\rho}\nede^*$. Here and henceforth an asterisk denotes evaluation at decay time $t_*$.

Within this general description, we can capture the presence of the trigger in terms of a function $q(t,\mathbf{x})$. It defines the transition time $t_*(\mathbf{x})$ through the implicit condition
\begin{align}\label{eq:q}
q(t_*(\mathbf{x}), \mathbf{x}) = q_* = \mathrm{const}\,.
\end{align}
For example, in cold and hybrid NEDE, which relies on a scalar field trigger $\phi$, we set $q=\phi $ and identify $q_*$ with the threshold value $\phi_*$ at which quantum tunneling becomes efficient or the field rolls over the watershed-like saddle, respectively. On the other hand, in Hot NEDE, we use the dark sector temperature $T_d$ as a trigger and accordingly set $q=T_d$.

The trigger mechanism has important implications for the perturbations within the NEDE fluid. Before the transition, we have $\delta \rho\nede = 0 $ because a cosmological constant does not support its own adiabatic perturbations (due to the constancy of $\bar{\rho}\nede$). After the transition, however,  $\bar{\rho}\nede$ is a quickly decaying fluid, which  will support its own fluctuations $\delta \rho\nede \neq 0$.
They are seeded at the transition time $t_*$ by the perturbations of the trigger, $\delta q_*(\mathbf{x}) \equiv \delta q(\bar{t}_*, \mathbf{x})$, where we decomposed
\begin{align} \label{eq:q_decompose}
q(t,\mathbf{x}) = \bar{q}(t) + \delta q (t, \mathbf{x}) \,.
\end{align}
These perturbations constitute dark sector acoustic oscillations and are a crucial part of the NEDE phenomenology. In particular, they lead to an excess decay of the Weyl potential, which balances the gravitational effect of an increased dark matter energy density (see the discussions in~\cite{Lin:2019qug,Niedermann:2020dwg,Vagnozzi:2021gjh}). They also make NEDE phenomenologically distinct from other early dark energy models.

The precise relation between $\delta q_*$ and $\delta \rho\nede^* \equiv \delta \rho\nede(\bar{t}_*)$ was derived in \cite{Niedermann:2020dwg} by  performing a fully covariant matching of the cosmological perturbation theory across the spacelike transition surface defined through Eq.~\ref{eq:q}.
Here, we will provide an equivalent but slightly more intuitive derivation of the same relation.   A spatial variation of Eq.~\ref{eq:q} yields
\begin{align}
 \left(\frac{\partial q_*}{\partial t_*} \, \frac{\partial t_*}{\partial \mathbf{x}}  +
\frac{\partial q_*}{\partial \mathbf{x}}  \right) \cdot d\mathbf{x} = 0\,,
\end{align}
which after substituting Eq.~\eqref{eq:q_decompose} becomes
\begin{align}\label{eq:delta_t_delta_q}
\delta t_*(\mathbf{x}) = - \frac{\delta q_*(\mathbf{x})}{\dot{\bar{q}}_*}\,,
\end{align}
at linear order in $\delta q$, thus relating the trigger variations $\delta q_*(\mathbf{x})$, with variations of the transition time $\delta t_*(\mathbf{x})$. This formula can be understood intuitively as follows; when the trigger's value increases, i.e.\ $\dot{\bar{q}} > 0$, then for a positive perturbation $\delta q(\mathbf{x})$ the threshold value $q=q_*$ will be reached slightly earlier, which indeed translates to a negative $\delta t_*(\mathbf{x})$.

Now, the perturbations in $\rho\nede$ arise because, due to the spatial dependence in Eq.~\ref{eq:delta_t_delta_q}, the fluid starts to decay at slightly different times at different positions in space. This can be described more formally by generalizing Eq.~\ref{eq:rho_nede} to
\begin{align}\label{eq:rho_nede_2}
\rho\nede(t,\mathbf{x}) \simeq \bar{\rho}\nede^* \exp{- 3 \int_{t_*(\mathbf{x})}^t  \mathrm{d} \tilde t  H \left[1 + w\nede(\tilde t)\right]}\,,
\end{align}
which is an expression for the full $\rho\nede(t,\mathbf{x})$, including perturbations, valid close to the transition, i.e. for $t \to t_*(\mathbf{x})$. The derivation of Eq.~\ref{eq:rho_nede_2} is simple: We can always choose coordinates for which the transition surface is perfectly flat and hence $\delta q_* = 0$ (or $\delta t_*=0$ equivalently). In this gauge $\delta \rho\nede(\bar{t}_*) = 0$ and thus the background formula in Eq.~\ref{eq:rho_nede} is an arbitrarily good approximation for the full $\rho\nede$ if evaluated sufficiently close to the transition surface (perturbations $\delta \rho\nede$ will be generated eventually due to gravitational sourcing). Transforming back to the original coordinates then yields  Eq.~\ref{eq:rho_nede_2}, where the spatial dependence of $\rho\nede(t,\mathbf{x})$ arises from $t_*(\mathbf{x})$ in the lower integration limit. Finally, perturbing Eq.~\ref{eq:rho_nede_2} and evaluating it at $t=t_*$ yields
\begin{subequations}\label{eq:matching}
\begin{align}
\frac{\delta \rho\nede^*}{\bar{\rho}_*} = -3 \left[1 + w\nede(t_*) \right] H_* \frac{\delta q_*}{\dot{\bar{q}}_*}\,,
\end{align}
where we used Eq.~\ref{eq:delta_t_delta_q} to substitute for  $\delta t_*$. This indeed agrees with the result obtained in~\cite{Niedermann:2020dwg}. There, a more exhaustive matching analysis also determines the initial velocity divergence $\theta\nede$, defined in momemtum space through the perturbed energy momentum tensor as $i k^i (\delta T\nede)^0_i \equiv (1 + w\nede)\bar{\rho}\nede \, a \, \theta\nede$ alongside $\delta \rho \nede \equiv - (\delta T\nede)_0^0$; explicitly,
\begin{align}
\theta\nede^* = \frac{k^2}{a_*} \frac{\delta q_*}{\dot{\bar{q}}_*}\,.
\end{align}
\end{subequations}
In synchronous gauge the metric takes the form
\begin{subequations}
\begin{align}\label{eq:sync}
ds^2 =  - dt^2  +a(t)^2 \left(\delta_{ij} + h_{ij} \right)dx^i dx^j \,,
\end{align}
where
\begin{align}
h_{ij} = \frac{k_i k_j}{k^2} h + \left(\frac{k_i k_j}{k^2} - \frac{1}{3} \delta_{ij}\right)6 \eta\,,
\end{align}
\end{subequations}
and $h=\delta^{ij} h_{ij}$. The evolution of $\delta\nede = \delta \rho\nede / \bar{\rho}\nede$ and $\theta\nede$ is then controlled by the two perturbation equations ~\cite{Ma:1995ey} (using cosmological time and neglecting spatial curvature)
\begin{subequations}\label{eq:pert}
\begin{align}
	\dot\delta\nede &= -3 \left[c_s^2 - w\nede(t)\right] H \delta\nede -\left[1\! +\! w\nede(t)\right]\left( \frac{\theta\nede}{a}\! + \!\frac{\dot h}{2}\right) \nonumber \\
	& \hspace{4.5cm}- 9 \left[1\! +\! w\nede(t)\right]\left[c_s^2 -c_a^2\right]H^2 \frac{\theta\nede}{a}\frac{a^2}{k^2}\,,\\[7pt]
	\dot\theta\nede  &= -\left(1-3c_s^2\right) H \theta\nede + c_s^2\frac{k^2}{a}\frac{\delta\nede}{1+w\nede(t)}\,, %- \frac{k^2}{a}\sigma\nede
\end{align}
\end{subequations}
where $c_s$ is the effective sound speed in the fluid's rest-frame and the adiabatic sound speed is fixed in terms of $w\nede$,
\begin{equation}
	c_a^2 = w\nede(t) - \frac{1}{3} \frac{\dot w\nede(t)}{1 + w\nede(t)} \frac{1}{H} \,.
	\label{eq:adiabatic_sound_speed}
\end{equation}
For simplicity, we assumed vanishing shear stress, $\sigma\nede=0$, which is a fully self-consistent choice. The generalized equations in the presence of $\sigma\nede$ and a phenomenological discussion thereof can be found in~\cite{Niedermann:2020dwg}.

The system up this point provides a very general phenomenological framework applicable to different microscopic implementations. From here on, we will focus on Cold NEDE, discussed in detail in~\cite{Niedermann:2020dwg}. In this case, the trigger $q(t,\mathbf{x})$ is identified with a subdominant scalar field $\phi(t,\mathbf{x}) = \bar{\phi}(t) + \delta \phi(t,\mathbf{x})$. Before the transition it satisfies the background equation
\begin{align}\label{eq:eom_clock}
\ddot{\bar{\phi}} + 3 H \dot {\bar{\phi}} + m^2 \bar{\phi} =0\,.
\end{align}
In particular, it is frozen for $H \gg m$ due to the Hubble friction term. Once $ H \lesssim m$, the field drops out of slow roll and starts to oscillate. For a radiation dominated universe, the solution is $\bar{\phi}(t) = \sqrt{2} \, \Gamma(5/4) \, \bar{\phi}_\text{ini} \, (H/m)^{1/4} \,  J_{1/4}\left(\frac{m}{2 H}\right)$, where $J_{1/4}$ and $\Gamma$ are the Bessel function of the first kind and the Gamma function, respectively. In particular, $\bar{\phi}/\bar{\phi}_\mathrm{ini} \to 0$ as $H/m \to 0.18$. Cold NEDE uses the idea that $\phi$ triggers a first order phase transition in another field $\psi$ as it approaches zero for the first time. This is  achieved through
the zero-temperature potential (assuming canonically normalized fields $\psi$ and $\phi$)
\begin{equation}
	V(\psi,\phi) = \frac{\lambda}{4}\psi^4 + \frac{1}{2} M^2\psi^2 - \frac{1}{3}\alpha M \psi^3 + \frac{1}{2} m^2\phi^2 + \frac{1}{2}\tilde{\lambda}\phi^2 \psi^2,
	\label{eq:potential}
\end{equation}
where $\alpha$ and $\lambda$ are positive and dimensionless coefficients of order unity.\footnote{A similar model has been considered before in an inflationary context~\cite{Linde:1990gz,Adams:1990ds,Copeland:1994vg}.} This potential comes with two mass scales: the heavy scale $M \sim \mathrm{eV}$ is setting the scale of NEDE and the ultralight scale $m \sim 10^{-27} \mathrm{eV}$ makes sure that the trigger field starts to roll around matter-radiation equality.  We further require $\alpha^2 > 4\lambda$ to ensure the right vacuum structure. Radiative stability of the trigger sector further demands~$\tilde \lambda < \mathcal{O}(1) \times 10^3\, m^2/M^2 \ll 1$. Initially the system sits at $(\psi,\phi) = (0,\phi_\mathrm{ini})$ and  is prevented from tunneling to its true minimum located at $( \psi, \phi)_\text{t.v.} =  \left(\frac{M}{2 \lambda}\,[\alpha + \sqrt{\alpha^2-4 \lambda }\,],0\right)$ by a huge potential barrier arising from the last term in Eq.~\ref{eq:potential}. This barrier is eventually removed as $\phi$ starts to evolve towards smaller values. Tunneling becomes efficient when $\phi$ reaches a threshold value $\phi_* \ll \phi_\mathrm{ini}$ corresponding to\footnote{Note that in the Cold NEDE model $H/m$ is not a free parameter but theoretically predicted to be $H/m \approx 0.2$. In \cite{Niedermann:2019olb} we checked this prediction by treating it as a free parameter and found, as an experimental vindication of the trigger mechanism, that indeed the value $H/m \approx 0.2$ is also preferred by the data.} $H/m \approx 0.2$ (recall that the first zero crossing occurs as $H/m \to 0.18$, which corresponds to the removal of the last term in Eq.~\ref{eq:potential} and hence maximal tunneling probability). Within this model, we define NEDE as the energy that is released in the phase transition,
\begin{equation}
	\bar{\rho}^*\nede \,=\, V(0,\phi_*) - V(\psi^{\rm t.v.},0).
\end{equation}
Having this type of triggered first-order phase transition achieves different things:
\begin{itemize}
\item The phase transition happens late enough in the expansion history to have an impact on the sound horizon $r_s$ (without the trigger bubble percolation would happen much earlier when $H \sim \mathrm{eV}$ without any noticeable effect on $r_s$).
\item The transition is completed on very short time scales (compared to $1/H$).  This avoids that bubbles grow too large, which would lead to additional (and potentially threatening) anisotropies in the CMB. Moreover, it makes sure that we can describe the colliding bubble wall condensate as a homogenous and isotropic fluid on large scales, corresponding to an evolution as in Eq.~\ref{eq:rho_nede_2}. It similarly justifies neglecting shear stress perturbations, $\sigma\nede = 0$, on these scales.

\item The trigger field $\phi$ carries adiabatic fluctuations that seed acoustic oscillations in the decaying NEDE fluid, which, as we pointed out before, are a vital part of the NEDE phenomenology.
\end{itemize}
After the phase transition we are left with a large number of vacuum bubbles that expand and quickly start to collide with each other. On small scales, this field condensate is dominated by anisotropic stress. A crucial assumption then is that this manifests itself on large scale as an excess pressure that leads to a decay quicker than radiation, i.e. $w\nede(t > t_*) > 1/3$. Eventually, the condensate is expected to be converted into scalar field radiation and gravitational waves, corresponding to $w\nede \to 1/3$ (depending on the implementation other decay channels are possible too). As cosmological observables are only sensitive to $\rho\nede$ when it is most dominant around the phase transition, we will approximate it as a constant, i.e.~$w\nede(t > t_*) \approx w\nede^* = \mathrm{const}$.

Cold NEDE has been implemented in the Boltzmann code \texttt{TriggerCLASS}~\cite{Niedermann:2020dwg}\footnote{https://github.com/flo1984/TriggerCLASS}, which builds on the Cosmic Linear Anisotropic Solving System \texttt{CLASS}~\cite{Blas:2011rf}. It evolves the trigger field $\bar{\phi}$ with Eq.~\ref{eq:eom_clock} alongside its adiabatic perturbations $\delta \phi$, governed by
\begin{align}
\delta \ddot \phi  + 2 H \delta \dot \phi + (k^2/a^2 +  m^2) \delta \phi & = -\frac{1}{2} \dot h \dot{\phi}  \,. \label{eq:pert_phi}
\end{align}
The matching Eqs.~\ref{eq:matching} at $t=t_*$ are then used to initialize the perturbations in the decaying NEDE fluid ($\delta\nede$ and $\theta\nede$). Their subsequent evolution is described in terms of the system of Eqs.~\ref{eq:pert}.  The base model relies on the following choices:  $(w\nede^*,c_s^2, H_*/m ) = (2/3,2/3, 0.2)$. Generalizations thereof are considered in~\cite{Niedermann:2020dwg}. The remaining model parameters are (i) the fraction of NEDE $f\nede$ and (ii) the mass of the trigger field $m$. The latter determines the redshift $z_*$ of the phase transition (or $t_*$ equivalently) and can be inferred from the numerical background evolution. An approximation is given by
\begin{equation}
	z_* \approx 2 \times 10^4 (1 - f\nede)^{1/4} \left(\frac{3.4}{g^*_\mathrm{rel,vis}}\right)^{1/4} \left(\frac{H_*/m}{0.2}\right) 10^{0.5\log_{10}(m/m_0)-1.75},
\end{equation}
where $g^*_\mathrm{rel,vis}$ is the effective number of relativistic degrees of freedom in the visible sector.

\section{Data analysis and results}
\label{sec:mcmcAnalysis}
In this section, we perform the cosmological parameter extraction. We first establish our data pipeline in Sec.~\ref{subsec:freeEOS} and then discuss the inclusion of ACT, SPT and BICEP data in sections~\ref{subsec:nedeACT}, \ref{subsec:nedeSPT} and \ref{subsec:bicep}, respectively.

\subsection{MCMC Analysis}
\label{subsec:freeEOS}

For simulating the NEDE model, we employ the Boltzmann code \texttt{TriggerCLASS} (v.~4.0). Our MCMC analysis samples the six standard $\Lambda$CDM parameters (with flat priors and usual ranges): the baryon density $\Omega_\mathrm{b} h^2$, the cold dark matter density $\Omega_\mathrm{c} h^2$, the Hubble parameter $H_0$, the amplitude $\log_{10}(10^{10}A_s)$ and scalar tilt $n_s$ of the primordial spectrum, the reionization depth $\tau_\mathrm{reio}$ and where applicable the tensor-to-scalar ratio $r$ at a pivot scale of $k_*=0.05$ Mpc${}^{-1}$. The neutrino sector is assumed to contain two massless and one massive species with $M_\nu = 3.046$. The $\Lambda$CDM parameter set is supplemented by the following cold NEDE parameters: the fraction of NEDE $f\nede$, the trigger mass $\log_{10}(m_\phi)$ (where $m_\phi \equiv m$ is measured in units of $1/\mathrm{Mpc}$) and the equation of state of the NEDE fluid after the phase transition $w^*\nede$. We impose the respective prior ranges  $0 < f\nede < 0.3$,  $1.0  < \log_{10}(m_\phi) < 3.0$ and $ 1.0  < w^*\nede < 3.0 $. The emergence of slight bimodalities within this 3-parameter NEDE model will motivate us to fix\footnote{This bimodality was also observed previously by \cite{Poulin:2021bjr}, who treated it by exploring two different regimes for the trigger mass.} $w^*\nede=2/3$ (we drop the asterisk henceforth for notational convenience) and study a 2-parameter NEDE model instead. In addition, we assume the rest-frame sound speed $c_s^2 $ to equal the adiabatic sound speed defined in \eqref{eq:adiabatic_sound_speed}, thus $c_s^2 = w\nede = \mathrm{const}$. The trigger parameter is set to be $H_*/m_\phi=0.2$, in accordance with a scalar field trigger that drops out of slow-roll as discussed in Sec.~\ref{sec:theModel}.

For the MCMC sampling we use \texttt{Cobaya}~\cite{Torrado:2020dgo} and produce chains until the Gelman-Rubin criterion~\cite{Gelman:1992zz} satisfies $R-1<0.01$, if not stated otherwise. For each run eight independent Markov chains are computed and can be publicly accessed \footnote{\url{https://drive.google.com/drive/folders/1h_jwj1tFX1tcliToSLK0n891y9NHy_7q?usp=sharing}}.
Our combined analysis uses as a baseline the following likelihoods:
\begin{itemize}
	\item Planck 2018 TT,TE,EE+lowE and lensing, with all nuisance parameters~\cite{Planck:2019nip}, (using v.~15.0 of \texttt{clik}\footnote{\url{https://github.com/benabed/clik}} ),
	\item BAO: 6dF 2011 + SDSS DR7\&12 \cite{Beutler:2011hx, Ross:2014qpa, BOSS:2016wmc},
	\item Compilation of spectroscopically-confirmed SNe Ia, Pantheon \cite{Pan-STARRS1:2017jku},
	\item BBN: PArthENoPE \cite{Pisanti:2007hk}.
\end{itemize}
We will join these base datasets with different combinations of the following datasets:
\begin{itemize}
	\item \textbf{ACT:} The ACTpol DR4 power spectrum likelihood \cite{ACT:2020frw, ACT:2020gnv} \footnote{\url{https://github.com/ACTCollaboration/pyactlike}}. The data consists of measurements of the temperature and polarization CMB spectrum. It corresponds to the survey made between 2013 and 2016 at frequencies of 98 and 150 GHz and includes multipoles ranging from 600.5 to 7525.5 for the TT spectra and from 350.5 to 7525.5 for the TE/EE spectra. In order to avoid double counting and known tensions between Planck and ACT data, we follow the suggestion by the ACT collaboration and only include multipoles with  $\ell>1800$ when combined with the Planck likelihood.

	\item \textbf{S$\mathbf{H_0}$ES:} We use the values reported in \cite{Riess:2021jrx} and impose the following Gaussian prior on $H_0$:
	\begin{equation}
		\chi^2_{H_0} \equiv -2\log{\cal L}_{H_0} = \left(\frac{H_0 - 73.04\;\text{km/s/Mpc}}{1.04\;\text{km/s/Mpc}}\right)^2.
	\end{equation}

	\item  \textbf{SPT:} The SPT3G Y1\cite{SPT-3G:2021eoc} likelihood. The data corresponds to a solid angle of 1500 deg${}^2$ and includes binned values for the EE and TE power spectra for the multipole range $300\leq \ell < 3000$. It is reported by the SPT Collaboration that their best-fit parameter values are compatible with $\Lambda$CDM and their value of $H_0$ is compatible with the one deduced from Planck 2018.

	\item \textbf{BICEP:} The BICEP/Keck program consists of several millimeter wave receivers able to measure the B-mode components of the CMB polarization spectrum. An initial set of observations was made at 150 GHz over two years and has been supplemented in the following years starting in 2010. After a previous data release, aimed at collecting data around the 95 GHz band in 2014 and following one at 220 GHz in 2015, a more recent set of observations was made public. These data contains the observations done in 2016, 2017 and 2018 at 220GHz with which the tensor-to-scalar ratio, $r$, can be constrained even further. The data release, BICEP18, used in the present analyses covers an effective area of 600 square degrees at 95 GHz and 400 square degrees at 150 \& 220 GHz and gives information at the multipole range $20<\ell<330$ \cite{BICEP:2021xfz}.

\end{itemize}
As a first criterion to evaluate the model's prospects, we compute the change in the effective chi-square, $\chi^2 = -2 \ln \mathcal{L}$ with $\mathcal{L}$ being the model's likelihood evaluated for the best-fit values, relative to the $\Lambda$CDM concordance model. We also report for each of the MCMC runs performed the absolute $\chi^2$ values for each individual likelihood as well as for the total run. The remaining statistical discrepancy with respect to S$H_0$ES, assuming Gaussian posteriors, can also be found in the corresponding tables.

Additionally, given that the Hubble tension is solely driven by the S$H_0$ES measurement among the datasets we used, a second useful criterion is that of the tension difference of the maximum a posteriori, $Q_{\rm dmap}$, advocated for in~\cite{Raveri:2018wln,Schoneberg:2021qvd}. It quantifies how much the quality of a given fit changes when the S$H_0$ES likelihood is included. For a given dataset combination $\mathcal{M}$, it is defined as
\begin{equation}
Q_{\rm dmap} \equiv \sqrt{\chi^2_{\mathcal{M}+\text{S}H_0\text{ES}}-\chi^2_{\mathcal{M}}},
\end{equation}
and is expected  to reduce to the normal tension measure if the posteriors are Gaussian.  Most crucially, it can be applied in situations where the posteriors are non-Gaussian as in the case of NEDE.
Intuitively, it tries to capture the idea that the “true” model should see the “true” physics even without S$H_0$ES and hence minimize $Q_\mathrm{dmap}$. In addition, we will also state the conventional Gaussian tension.

\subsection{Study of NEDE including ACT}
\label{subsec:nedeACT}

Here, we will study how NEDE reacts to including additional CMB temperature and polarization data from ACT. First, in Sec.~\ref{sec:Act_bimodality}, we will find that including ACT data leads to a bi-modality when fitting to the 3-parameter NEDE model. We will therefore propose a prescription to isolate the dominant mode. This, in turn, motivates our 2-parameter baseline model that will be discussed in Sec.~\ref{sec:Act_main}.

\subsubsection{Establishing the Baseline Model}\label{sec:Act_bimodality}

\begin{figure}[htbp]
	\centering
	\includegraphics[clip, width=0.95\textwidth]{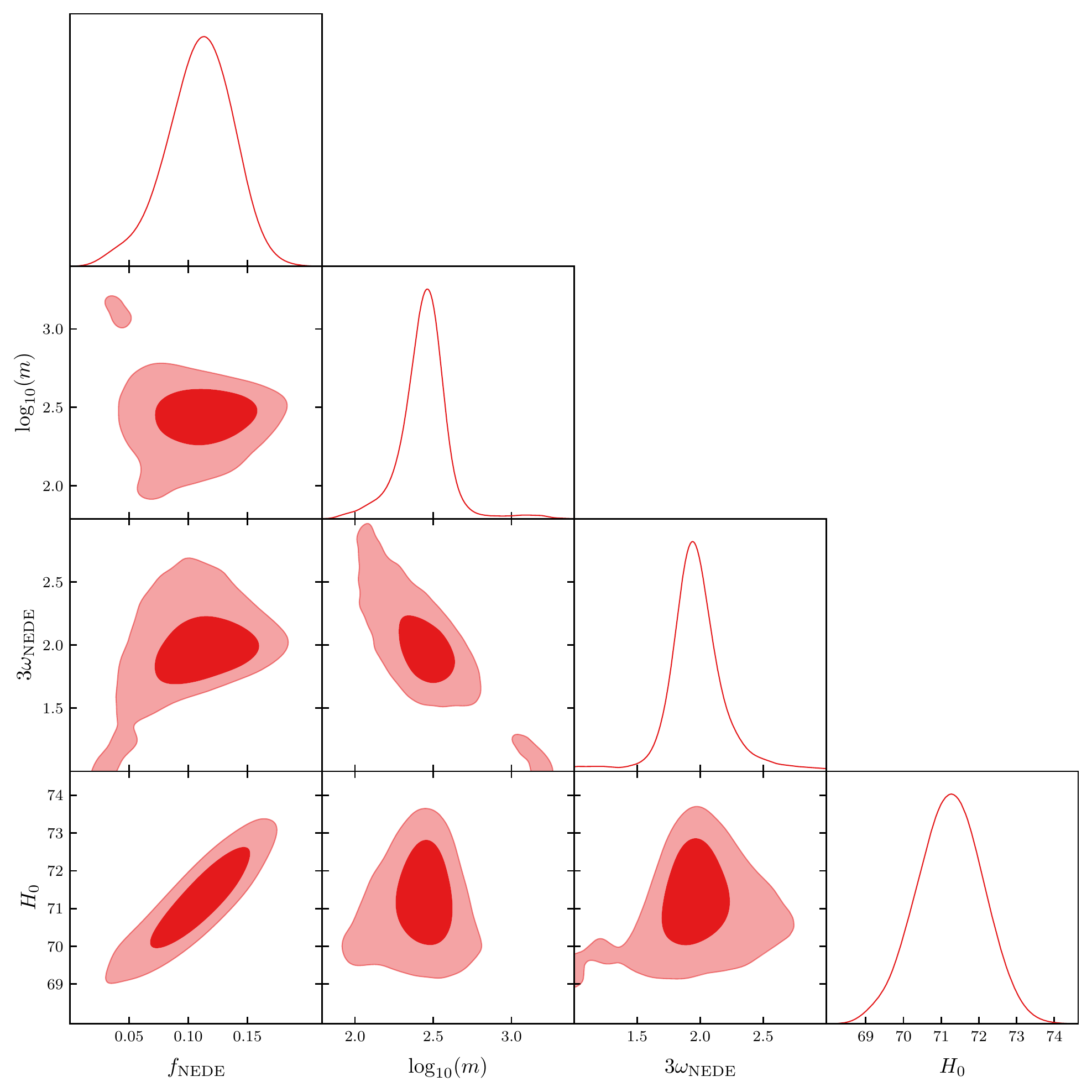}
%	\begin{minipage}[b][8.5cm][c]{.35\textwidth}
%		\includegraphics[clip, width=\textwidth]{H0vsMwEOS}
%	\end{minipage}
	\caption{(Left) Covariances and posteriors for joint analysis of the base datasets with ACT and S$H_0$ES. There is a slight bi-modality: besides the dominant mode that shows typical NEDE behavior, there is weak evidence for a high-mass / low-$w\nede$ mode. Later we choose the dominant mode by fixing $w\nede = 2/3$.}
	\label{fig:smallTriangle}
\end{figure}

We first consider the 3-parameter NEDE model where $w\nede$ is allowed to vary along with $f\nede$ and $\log_{10}(m_\phi)$. This is a joint analysis using the base datasets together with ACT and S$H_0$ES.  We present the triangle plot for this run for the NEDE parameters in Fig.~\ref{fig:smallTriangle}. The corresponding mean and best-fit values  are reported in Tab.~\ref{tab:mcmcACTfreeEOS}. Further derived parameters can be found in the Appendix in Tab.~\ref{tab:derivedFreeEOS}

Quite remarkably, when compared with the corresponding $\Lambda$CDM fit in Tab.~\ref{tab:bestfitLCDM}, there is a $\chi^2$ improvement of around $20$ units, and a nonzero fraction of the NEDE fluid ($f\nede > 0$) is favored with more than 3.5$\sigma$ preference. Moreover, we obtain $H_0 = 71.2 \pm 0.9$ km/s/Mpc (and best-fit $71.50$ km/s/Mpc), which is in agreement with previous studies and confirms that the NEDE model is able to reduce the Hubble tension significantly.

These initial results, albeit reproducing the evidence for NEDE seen in previous analyses without ACT~\cite{Niedermann:2020dwg,Schoneberg:2021qvd}, have to be taken with a grain of salt. The reason is that the posteriors for $w\nede$ and $\log_{10}(m_\phi)$ show a slight bi-modality (see Fig~\ref{fig:smallTriangle}), giving (weak) statistical support for the presence of an additional high-mass / low-$w\nede$ mode at $w\nede\approx 1/3$ and $\log_{10}(m_\phi) \approx 3$ (corresponding to $z_* \approx 9000$). This questions the reliability of a conventional parameter extraction which generically assumes single-mode posteriors. As we will see, it also leads to an understatement of the model's performance as a solution to the Hubble tension. Nevertheless, the observed bi-modality does not prevent the MCMC chains from converging using the standard sampling, and the dominant mode seems to be robust enough. For models with several equally strong modes, a nested sampler may be considered, but we deem this is not necessary for NEDE at this stage. We will, however, seek to confirm these conclusions in a future publication making use of a complementary approach using profile likelihoods as a way to deal with bimodalities~(see \cite{Planck:2013nga} and for recent work in the context of EDE also~\cite{Herold:2021ksg}). Here, we will take another less computationally heavy path and fix $w\nede = 2/3$.
It can be seen from Fig.~\ref{fig:smallTriangle} that this value corresponds to the dominant mode, in agreement with previous findings in the literature \cite{Poulin:2021bjr}. Moreover, for this mode, also the other model parameters remain consistent with the values obtained in a previous analysis without ACT. In particular, besides the shift in $H_0$, we find an increase in $\omega_\mathrm{c}$, $\sigma_8$ and $n_s$ (with relevance for inflationary model building), which is typical for models that feature an early dark energy component.

\begin{table}[htbp]
	\centering
	\renewcommand{\arraystretch}{1.3}
	\setlength\tabcolsep{0pt}
	\fontsize{8}{11}\selectfont    
\begin{tabular}{|>{\centering\arraybackslash}p{4cm}|>{\centering\arraybackslash}p{3.2cm}|}
	\hline
	\hspace*{.5cm}{\centering Name}\hspace*{.5cm}                                                                           & \hspace*{.3cm}Best-fit\newline Mean$_{\rm Lower}^{\rm Upper}$ \\ \hline\hline
	\multirow{2}{1.5cm}{\centering $\Omega_\mathrm{b} h^2$ }                                                                & $0.023$                                                       \\
	                                                                                                                        & $0.022693\pm 0.000185$                                        \\ \hline
	\multirow{2}{1.5cm}{\centering $\Omega_\mathrm{c} h^2$ }                                                                & $0.130$                                                       \\
	                                                                                                                        & $0.12840\pm 0.00303$                                          \\ \hline
	\multirow{2}{1.5cm}{\centering $H_0$ }                                                                                  & $71.502$                                                      \\
	                                                                                                                        & $71.230\pm 0.890$                                             \\ \hline
	\multirow{2}{1.5cm}{\centering $\log(10^{10} A_\mathrm{s})$\hspace*{.5cm} }                                             & $3.083$                                                       \\
	                                                                                                                        & $3.0702\pm 0.0141$                                            \\ \hline
	\multirow{2}{1.5cm}{\centering $n_\mathrm{s}$ }                                                                         & $0.988$                                                       \\
	                                                                                                                        & $0.98855^{+0.00645}_{-0.00568}$                               \\ \hline
	\multirow{2}{1.5cm}{\centering $\tau_\mathrm{reio}$ }                                                                   & $0.054$                                                       \\
	                                                                                                                        & $0.05607\pm 0.00715$                                          \\ \hline
	\rowcolor{gray!30}                                                                                                      & $0.120$                                                       \\
	\rowcolor{gray!30} \multirow{-2}{1.5cm}{\centering $f\nede$ }                                                           & $0.1093^{+0.0311}_{-0.0246}$                                  \\ \hline
	\rowcolor{gray!30}                                                                                                      & $2.459$                                                       \\
	\rowcolor{gray!30} \multirow{-2}{1.5cm}{\centering \hspace*{-.3cm}$\log_{10}(m_\phi/\textrm{Mpc}^{-1})$\hspace*{.5cm} } & $2.444^{+0.131}_{-0.112}$                                     \\ \hline
	\rowcolor{gray!30}                                                                                                      & $1.902$                                                       \\
	\rowcolor{gray!30} \multirow{-2}{1.5cm}{\centering $3\omega\nede$ }                                                     & $1.974^{+0.157}_{-0.202}$                                     \\ \hline
	\rowcolor{gray!30}                                                                                                      & $287.950$                                                     \\
	\rowcolor{gray!30} \multirow{-2}{1.5cm}{\centering $m_\phi$ [Mpc${}^{-1}$] }                                            & $302^{+50}_{-100}$                                            \\ \hline
	\rowcolor{gray!30}                                                                                                      & $4290.739$                                                    \\
	\rowcolor{gray!30} \multirow{-2}{1.5cm}{\centering $z_\mathrm{decay}$ }                                                 & $4322^{+600}_{-800}$                                          \\ \hline
	$\sum \chi^2$                                                                                                           & $4044.31$                                                     \\ \hline
	$\Delta \chi^2$                                                                                                         & $-20.63$                                                      \\ \hline
\end{tabular}
\caption{Posterior means, $1\sigma$ confidence intervals and best-fit values for the MCMC simulation using the base datasets + ACT + S$H_0$ES for the 3-parameter NEDE model, allowing sampling over $w\nede$. Shaded in gray are the NEDE-specific parameters together with the derived parameters $m_\phi$ and $z_{*}$. NEDE is preferred over $\Lambda$CDM with a significance of more than $3.5 \sigma$.}
\label{tab:mcmcACTfreeEOS}
\end{table}

\subsubsection{Results and Discussion}\label{sec:Act_main}
\begin{figure}[htbp]
	\centering
	\includegraphics[clip, width=.6\textwidth]{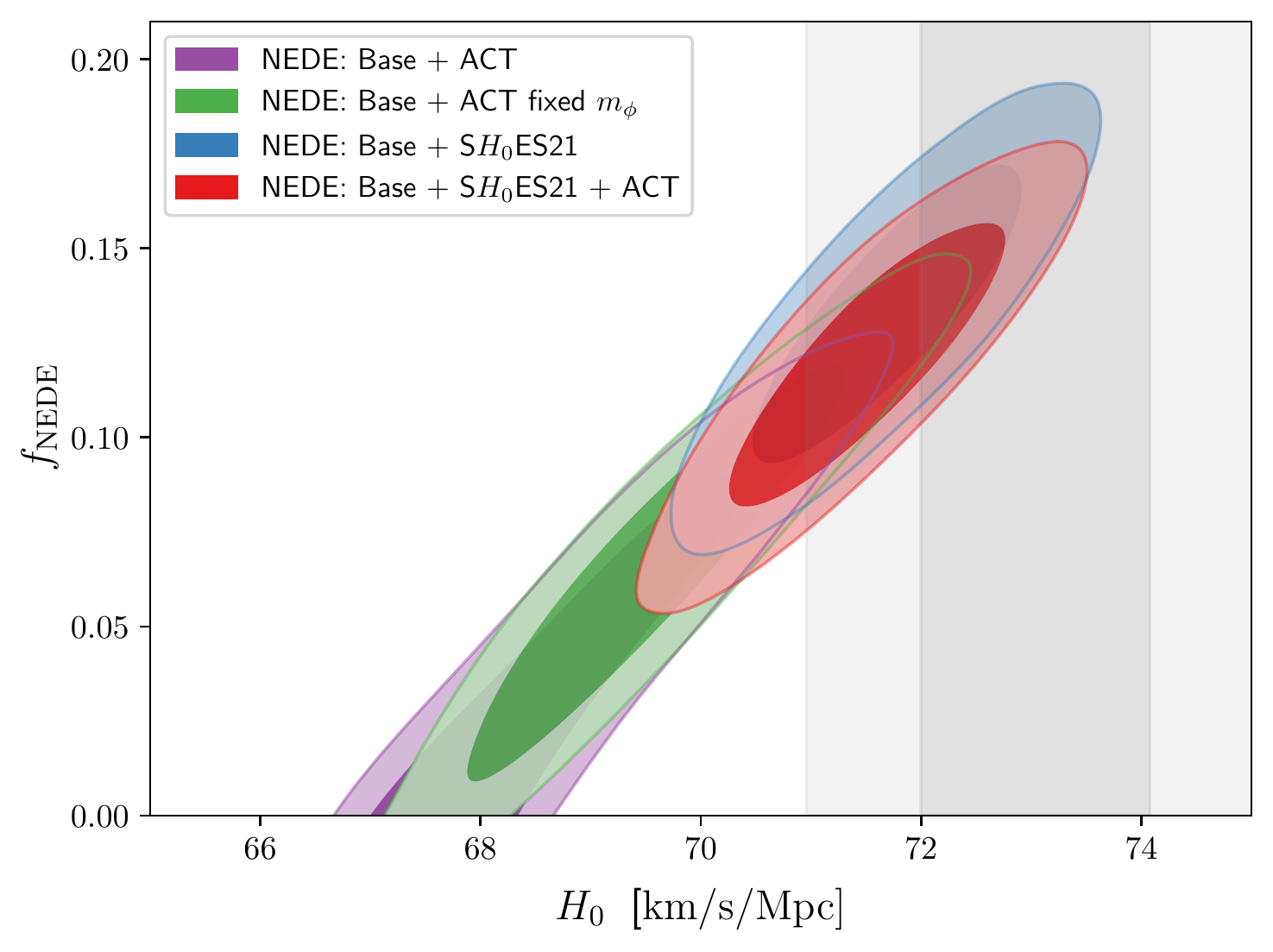}
	\caption{Covariance between $f\nede$ and $H_0$ for different dataset combinations featuring the new ACT dataset. The NEDE model is subject to a fixed equation of state $w\nede=2/3$.}
	\label{fig:fNEDEvsH0ACT}
\end{figure}

We present the posterior means with their $1\sigma$ confidence intervals and best-fit values for the 2-parameter NEDE model (with $w\nede=2/3$) in Tab.~\ref{tab:bestfitACT}, along with their corresponding $\Delta \chi^2$ and $Q_{\rm dmap}$ tension. Our main findings with regards to the Hubble tension are summarized in Fig.~\ref{fig:fNEDEvsH0ACT}, depicting the $f\nede$ vs $H_0$ contours. For completeness, the covariance contours for the six $\Lambda$CDM parameters can be found in Fig.~\ref{fig:triangleAltACT} in the appendix. There, we also state the values of the different derived parameters (see Tab.~\ref{tab:derivedPosteriorsACT}), and display the individual $\chi^2$ per dataset (see Tab.~\ref{tab:chiSqLCDM} and Tab.~\ref{tab:chiSqACT} for $\Lambda$CDM and NEDE, respectively).

As a first sanity check, we use the baseline dataset with and without S$H_0$ES to re-derive the results obtained in~\cite{Niedermann:2020dwg}, this time with the Cobaya pipeline as described above (the old analysis used MontePython). We find that the extracted parameters deviate less than  $0.5 \sigma$ from their previously derived counterparts. In particular, with S$H_0$ES included, there is a fit improvement over $\Lambda$CDM of $\Delta \chi^2 = -15.9 $, and we have a sizable fraction of NEDE ($f\nede  = 13.3 \pm 2.6 \%$) that is accompanied by an increased value $H_0 = 71.7 \pm 0.8\,\mathrm{km/s/Mpc}$, reducing the $Q_\mathrm{dmap}$ tension below $2 \sigma$ (down from $4.8 \sigma$ in $\Lambda$CDM) in agreement with the findings in~\cite{Schoneberg:2021qvd} (see also blue contour in Fig.~\ref{fig:fNEDEvsH0ACT}). We further observe a $2.5 \%$ increase in $n_s$ (and a milder increase in the spectral amplitude) when comparing with the corresponding $\Lambda$CDM run. As mentioned before in Sec.~\ref{sec:Act_main} this change is required to counter the gravitational effect of the dark sector acoustic oscillations that are typical for early dark energy models.

It is typically observed that without including S$H_0$ES the evidence for NEDE seems much smaller (although the data fit still improves by $\Delta \chi^2 = -1.8$).  By now, the reason for this has been investigated in detail by different authors~\cite{Niedermann:2020dwg, Murgia:2020ryi, Herold:2021ksg, Reeves:2022aoi}: when the sampling algorithm probes smaller values of $f\nede $  -- as expected for a run without the statistical pull of the S$H_0$ES dataset -- the other NEDE parameter $\log_{10}(m_\phi)$ (or $z_*$ equivalently) becomes less constrained (obviously, if there is only a small energy injection it does not matter when it happens). This corresponds to an increase in the overall sampling volume as  $f\nede \to 0 $  which makes the MCMC chain collect more data points in the low-$f\nede$ region. This can be diagnosed independently through a non-Gaussian posterior for $f\nede$, which rather generically suggests a strong and unphysical prior dependence.
\begin{table}[htbp]
	\begin{center}
		\renewcommand{\arraystretch}{1.3}
		\setlength\tabcolsep{0pt}
		\fontsize{8}{11}\selectfont
		\begin{tabular}{|>{\centering\arraybackslash}p{2.4cm}||>{\centering\arraybackslash}p{2.5cm}|>{\centering\arraybackslash}p{2.5cm}|>{\centering\arraybackslash}p{2.5cm}|>{\centering\arraybackslash}p{2.5cm}|>{\centering\arraybackslash}p{2.5cm}|}
			\hline
			\multirow{3}{1.5cm}{Parameter Name}                                                                 &                                                                                              \multicolumn{5}{c|}{NEDE fixed EOS}                                                                                               \\
			\hhline{~|----|-|}                                                                                  & \multirow{2}{1.5cm}{\centering Base} & \multirow{2}{1.5cm}{\centering+ACT} & \multirow{2}{1.5cm}{\centering+S$H_0$ES21} & \multirow{2}{1.5cm}{\centering+ACT +S$H_0$ES21} & \multirow{2}{1.5cm}{\centering+ACT fixed $m_\phi$} \\
			                                                                                                    &                                      &                                     &                                            &                                                 &                                                    \\ \hline\hline
			\multirow{2}{1.5cm}{\centering $\Omega_\mathrm{b} h^2$}                                             & $0.023$                              & $0.022$                             & $0.023$                                    & $0.023$                                         & $0.023$                                            \\
			                                                                                                    & $0.0226^{+0.00020}_{-0.00025}$       & $0.0225\pm 0.00019$                 & $0.0230\pm 0.00022$                        & $0.0227\pm 0.00017$                             & $0.0226\pm 0.00015$                                \\ \hline
			\multirow{2}{1.5cm}{\centering $\Omega_\mathrm{c} h^2$}                                             & $0.125$                              & $0.124$                             & $0.131$                                    & $0.131$                                         & $0.129$                                            \\
			                                                                                                    & $0.1244^{+0.0025}_{-0.0048}$         & $0.1230^{+0.0021}_{-0.0039}$        & $0.1308\pm 0.00301$                        & $0.1293\pm 0.00284$                             & $0.1251^{+0.0028}_{-0.0036}$                       \\ \hline
			\multirow{2}{1.5cm}{\centering $H_0$}                                                               & $69.444$                             & $69.015$                            & $71.759$                                   & $72.086$                                        & $70.955$                                           \\
			                                                                                                    & $69.306^{+0.883}_{-1.48}$            & $68.945^{+0.746}_{-1.34}$           & $71.698\pm 0.801$                          & $71.485\pm 0.822$                               & $69.679^{+0.940}_{-1.25}$                          \\ \hline
			\multirow{2}{1.5cm}{\centering $\log(10^{10} A_s)\;\;$}                                             & $3.047$                              & $3.064$                             & $3.065$                                    & $3.084$                                         & $3.078$                                            \\
			                                                                                                    & $3.056\pm 0.0155$                    & $3.0577\pm 0.0141$                  & $3.0688\pm 0.0146$                         & $3.0710\pm 0.0144$                              & $3.0620\pm 0.0148$                                 \\ \hline
			\multirow{2}{1.5cm}{\centering $n_s$}                                                               & $0.978$                              & $0.976$                             & $0.991$                                    & $0.995$                                         & $0.986$                                            \\
			                                                                                                    & $0.9765^{+0.0069}_{-0.0093}$         & $0.9757^{+0.0066}_{-0.0084}$        & $0.9910\pm 0.0057$                         & $0.9905^{+0.0061}_{-0.0054}$                    & $0.9810^{+0.0063}_{-0.0075}$                       \\ \hline
			\multirow{2}{1.5cm}{\centering $\tau_{\rm reio}$}                                                   & $0.055$                              & $0.050$                             & $0.054$                                    & $0.059$                                         & $0.053$                                            \\
			                                                                                                    & $0.0564\pm 0.0073$                   & $0.0547^{+0.0064}_{-0.0073}$        & $0.0575\pm 0.0074$                         & $0.0559\pm 0.0073$                              & $0.0548\pm 0.0072$                                 \\ \hline\hline
			\rowcolor{gray!30}                                                                                  & $0.067$                              & $0.053$                             & $0.135$                                    & $0.139$                                         & $0.110$                                            \\
			\rowcolor{gray!30}\multirow{-2}{1.5cm}{ \centering $f\nede$}                                        & $< 0.132$                            & $< 0.107$                           & $0.1330\pm 0.0255$                         & $0.119^{+0.0265}_{-0.0229}$                     & $0.0690^{+0.0326}_{-0.0370}$                       \\ \hline
			\rowcolor{gray!30}                                                                                  & $2.543$                              & $2.458$                             & $2.583$                                    & $2.468$                                         & $2.4583$                                           \\
			\rowcolor{gray!30}\multirow{-2.7}{1.5cm}{ \centering \hspace*{-.3cm}$\log_{10}\left(\dfrac{m_\phi}{\textrm{Mpc}^{-1}}\right)$}                           & $2.554\pm 0.286$                     & $2.313^{+0.427}_{-0.158}$           & $2.553\pm 0.100$                           & $2.445^{+0.105}_{-0.0800}$                      & --                                                 \\ \hline
			\rowcolor{gray!30}                                                                                  & $349.486$                            & $287.333$                           & $382.974$                                  & $293.758$                                       & $287.333$                                          \\
			\rowcolor{gray!30}\rowcolor{gray!30}\multirow{-2}{1.5cm}{\centering $m_\phi$ {\tiny[Mpc${}^{-1}$]}} & $< 1090$                             & $< 709$                             & $367^{+70}_{-90}$                          & $286^{+60}_{-60}$                               & --                                                 \\ \hline
			\rowcolor{gray!30}                                                                                  & $4881.454$                           & $4397.410$                          & $5007.738$                                 & $4306.963$                                      & $4301.678$                                         \\
			\rowcolor{gray!30}\rowcolor{gray!30}\multirow{-2}{1.5cm}{\centering $z_*$}                          & $5270^{+1000}_{-2000}$               & $4107^{+1000}_{-3000}$              & $4856^{+600}_{-600}$                       & $4247\pm 600$                                   & $4370^{+63}_{-52}$                                 \\ \hline\hline
			\hspace*{.1cm}	Total $\chi^2$                                                                       & 3804.26                              & 4040.56                             & 3806.74                                    & 4049.05                                         & 4039.26                                            \\ \hline
			\hspace*{.1cm}	$\Delta\chi^2$                                                                       & $-3.19$                              & \cellcolor{violet!40} $-1.82$       & \cellcolor{teal!40} $-23.32$               & \cellcolor{red!40} $-15.89$                     & \cellcolor{green!40} $-3.13$                       \\ \hline
			\hspace*{.1cm}	$H_0$ Tension                                                                        & $2.3\sigma$                          & $2.7\sigma$                         &                                   \multicolumn{2}{c|}{--}                                    & $2.2\sigma$                                        \\ \hline
			\hspace*{.1cm}	$Q_{\rm dmap}$                                                                       &                          \multicolumn{2}{c|}{--}                           & 1.57$\sigma$                               & 2.9$\sigma$                                     & --                                                 \\ \hline
		\end{tabular}
	\end{center}
	\caption{Best-fit results of the MCMC analysis involving the ACT data and pertinent likelihood combinations for reference. Colors correspond to the contours of Fig.~\ref{fig:fNEDEvsH0ACT}. If only a one-sided constraint exists at $95\%$CL, it is given in the table instead of the $68\%$ error.}
	\label{tab:bestfitACT}
\end{table}

We now turn to the ACT runs and first investigate them in the absence of the S$H_0$ES prior on $H_0$. Figure~\ref{fig:fNEDEvsH0ACT} shows how the NEDE fit to the baseline datasets, and ACT is still affected by the before-mentioned sampling issues near the lower values of $f\nede$, which is evident by the enlargement and widening of the 68\%  C.L. towards the bottom. To confirm that this is indeed a sampling issue and still be able to extract cosmological parameters, we perform one run where we fix the mass of the trigger field to its best-fit value as given by the unrestricted simulation to obtain the green contours. As a result, we indeed find that the green relative to the purple contour moves to larger values of $f\nede$ and, in fact, shows weak evidence for non-zero $f\nede$ (with $1\sigma$ significance). Even more important is the observation that NEDE makes the joint baseline and ACT datasets compatible with S$H_0$ES at the $95\%$ C.L., which motivates the subsequent inclusion of the S$H_0$ES prior on $H_0$.

Concerning the best-fit of the run including ACT (but without S$H_0$ES) there is some spurious improvement in $\chi^2$ for fixed mass compared to the one with unconstrained mass due NEDE featuring approximately degenerate minima or flat directions in that region of parameter space. We performed different minimization procedures, which led to very similar values of $\chi^2$ but still with some small variability in the cosmological parameters, however, all are compatible with the results for the posteriors.

In a nutshell, including ACT data in the presence of S$H_0$ES reproduces the baseline results qualitatively but leads to slightly weaker evidence for NEDE (compare the blue and red contour in Fig.~\ref{fig:fNEDEvsH0ACT}).
As usual, with the S$H_0$ES prior higher fractions of NEDE are favored; although, when also including ACT, $f\nede$ is lower by $\simeq 1.4\%$ as compared to the run without ACT. Specifically, we can see from Tab.~\ref{tab:bestfitACT} that the central value lies at around a 12\% fraction of NEDE and is favored over a null fraction by about 4.8$\sigma$. Without ACT we found a fraction of NEDE of around 13\% with a 5.2$\sigma$ preference over the null hypothesis.

We now highlight some of the remarkable features of the joint ACT analysis. The first observation relates to $\Lambda$CDM, where we see that when comparing corresponding datasets, NEDE achieves a better fit by decreasing the $\chi^2$ by more than 15 units for the full data combination. Secondly, compared with the baseline dataset, ACT seems to slightly reduce the mean NEDE fraction, although in a statistically insignificant way. Interestingly, the best-fit values remain high and unaffected by ACT, meaning that the distribution becomes less Gaussian. This best-fit value for the NEDE fraction comes at the price of a slight worsening of the fit compared to the same analysis without ACT.  Third, we can have a look at the second criterion, the $Q_{\rm dmap}$, which attempts to favor models which do not change much by the inclusion of local measurements.  In this regard, we see that ACT worsens the $Q_{\rm dmap}$ which is interpreted as ACT enhancing the pull that the S$H_0$ES likelihood has on the data.

\begin{figure}[htbp]
	\centering
	\includegraphics[clip, width=.95\textwidth]{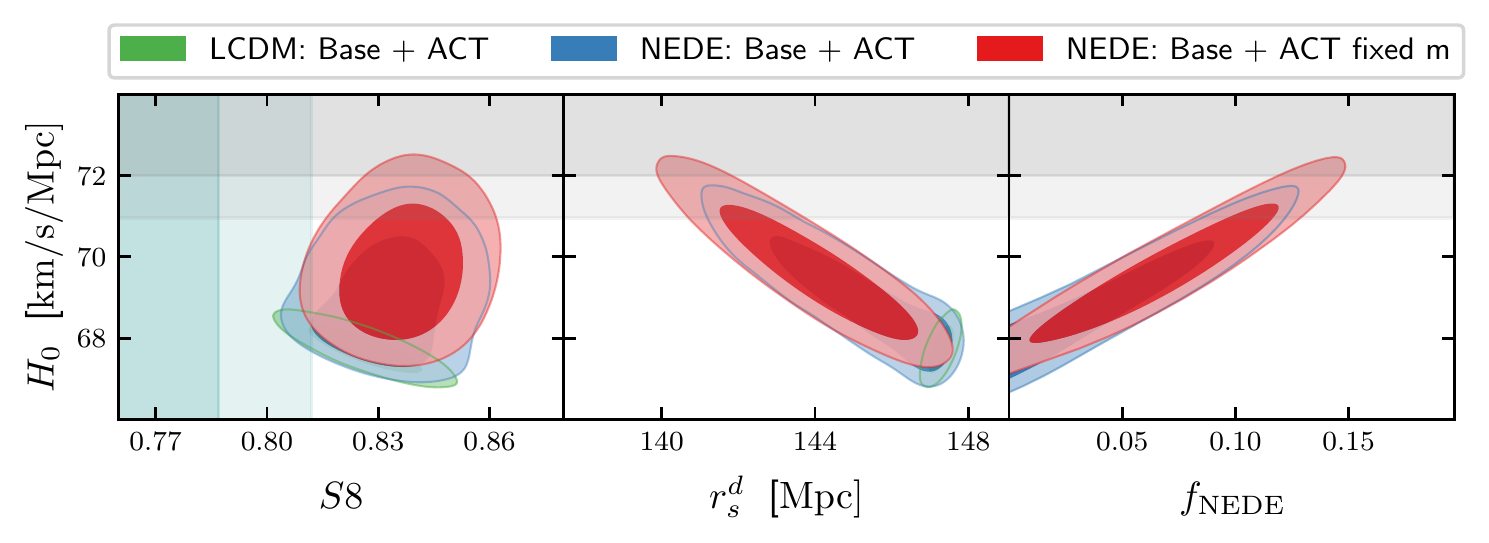}
	\caption{Plots showing the contours relating the Hubble parameter with $S8$, $r_s^{\rm d}$ and $f\nede$, while excluding the S$H_0$ES likelihood. The gray bands show the $1$ and $2\sigma$ contours coming from the latest S$H_0$ES report \cite{Riess:2021jrx}, while the teal bands correspond to the measurements from KiDS+VIKING and DES \cite{Joudaki:2019pmv}.}
	\label{fig:S8rsACT}
\end{figure}

As derived parameters we show the plots for $H_0$ against $S8$ and $r_s^{\rm d}$ in Fig.~\ref{fig:S8rsACT} and include the bands given by S$H_0$ES 2021\cite{Riess:2021jrx} and in the case of $S8$, also the bands from the cosmic shear combined analysis of KiDS+VIKING-450 and DES-Y1\cite{Joudaki:2019pmv}\footnote{The newer data releases point at similar constraints but have not been combined yet: $S8 = 0.759^{+0.024}_{-0.021}$~\cite{KiDS:2020suj} from KiDS - 1000 and $S8 = 0.772^{+0.018}_{-0.017}$~\cite{DES:2022qpf} from DES-Y3.}. The plots include the contours obtained for $\Lambda$CDM and clearly show how the NEDE model moves the contours towards the S$H_0$ES band, however, at the same time, the tension with the low-$z$ values for $S8$ is increased mildly. To be precise, NEDE  sits on the edge of the 2$\sigma$ region for $S8$ and does not improve on the existing tension. Some more comments on this are given in the conclusions.

\subsection{Study of NEDE including SPT}
\label{subsec:nedeSPT}

\begin{figure}[htbp]
	\centering
	\includegraphics[clip, width=.6\textwidth]{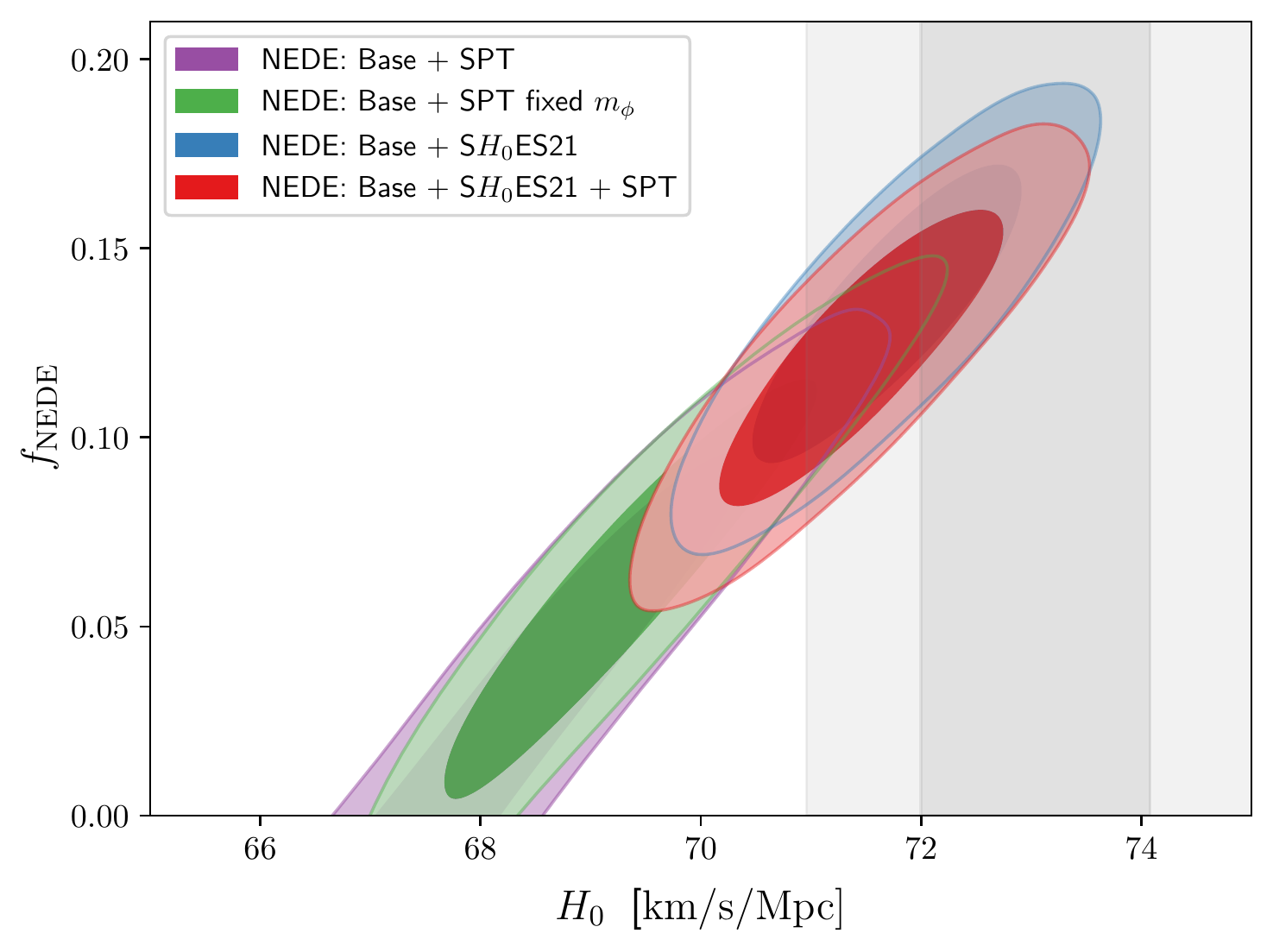}
	\caption{Plot of the covariance between $f\nede$ and $H_0$ for different dataset combinations featuring the new SPT dataset. The NEDE model is subject to a fixed equation of state $w\nede=2/3$.}
	\label{fig:fNEDEvsH0SPT}
\end{figure}

Here, we perform similar joint analyses for the case of the SPT likelihood. The results for the best-fit values, posterior means, and 1$\sigma$ confidence intervals are found in Table~\ref{tab:bestfitSPT}. A plot showing the covariance for $H_0$ against $f\nede$ is given in Fig.~\ref{fig:fNEDEvsH0SPT}. The full set of posteriors and marginalized distributions of the cosmological parameters shared with $\Lambda$CDM can be found in the triangle plot of Fig.~\ref{fig:triangleAltSPT}. We include the results for some derived parameters in Table~\ref{tab:derivedPosteriorsSPT} and the $\chi^2$ for the individual datasets for runs involving SPT in Tab.~\ref{tab:chiSqSPT}.
\begin{table}[htbp]
	\begin{center}
		\renewcommand{\arraystretch}{1.3}
		\setlength\tabcolsep{0pt}
		\fontsize{8}{11}\selectfont
	\begin{tabular}{|>{\centering\arraybackslash}p{2.4cm}|>{\centering\arraybackslash}p{2.5cm}|>{\centering\arraybackslash}p{2.5cm}|>{\centering\arraybackslash}p{2.5cm}|>{\centering\arraybackslash}p{2.5cm}|>{\centering\arraybackslash}p{2.5cm}|}
		\hline
		\multirow{3}{1.5cm}{Parameter Name}                                                                 &                                                                                              \multicolumn{5}{c|}{NEDE fixed EOS}                                                                                               \\
		\hhline{~|-----|}                                                                                   & \multirow{2}{1.5cm}{\centering Base} & \multirow{2}{1.5cm}{\centering+SPT} & \multirow{2}{1.5cm}{\centering+S$H_0$ES21} & \multirow{2}{1.5cm}{\centering+SPT +S$H_0$ES21} & \multirow{2}{1.5cm}{\centering+SPT fixed $m_\phi$} \\
		                                                                                                    &                                      &                                     &                                            &                                                 &                                                    \\ \hline\hline
		\multirow{2}{1.5cm}{\centering $\Omega_\mathrm{b} h^2$}                                             & $0.023$                              & $0.023$                             & $0.023$                                    & $0.023$                                         & $0.023$                                            \\
		                                                                                                    & $0.0227^{+0.00020}_{-0.00025}$       & $0.0225^{+0.00016}_{-0.00020}$      & $0.0230\pm 0.00022$                        & $0.0228^{+0.00017}_{-0.00020}$                  & $0.0226\pm 0.00016$                                \\ \hline
		\multirow{2}{1.5cm}{\centering $\Omega_\mathrm{c} h^2$}                                             & $0.125$                              & $0.125$                             & $0.131$                                    & $0.129$                                         & $0.125$                                            \\
		                                                                                                    & $0.1244^{+0.0025}_{-0.0048}$         & $0.1233^{+0.0021}_{-0.0039}$        & $0.1308\pm 0.0030$                         & $0.1293\pm 0.0030$                              & $0.1247^{+0.0027}_{-0.0037}$                       \\ \hline
		\multirow{2}{1.5cm}{\centering $H_0$}                                                               & $69.444$                             & $69.319$                            & $71.759$                                   & $71.769$                                        & $69.461$                                           \\
		                                                                                                    & $69.306^{+0.883}_{-1.48}$            & $69.019^{+0.781}_{-1.27}$           & $71.698\pm 0.801$                          & $71.431\pm 0.852$                               & $69.461^{+0.901}_{-1.26}$                          \\ \hline
		\multirow{2}{1.5cm}{\centering $\log(10^{10} A_s)\;\;$}                                             & $3.047$                              & $3.055$                             & $3.065$                                    & $3.071$                                         & $3.057$                                            \\
		                                                                                                    & $3.0560\pm 0.0155$                   & $3.0494\pm 0.0147$                  & $3.0688\pm 0.0146$                         & $3.0617\pm 0.0143$                              & $3.0513\pm 0.0147$                                 \\ \hline
		\multirow{2}{1.5cm}{\centering $n_s$}                                                               & $0.978$                              & $0.978$                             & $0.991$                                    & $0.990$                                         & $0.978$                                            \\
		                                                                                                    & $0.9765^{+0.0069}_{-0.0093}$         & $0.9751^{+0.0062}_{-0.0080}$        & $0.9909\pm 0.0057$                         & $0.9888\pm 0.0059$                              & $0.9784^{+0.0064}_{-0.0077}$                       \\ \hline
		\multirow{2}{1.5cm}{\centering $\tau_{\rm reio}$}                                                   & $0.055$                              & $0.055$                             & $0.054$                                    & $0.061$                                         & $0.057$                                            \\
		                                                                                                    & $0.0564\pm 0.0073$                   & $0.0544\pm 0.0072$                  & $0.0575\pm 0.0074$                         & $0.0555\pm 0.0071$                              & $0.0542\pm 0.0070$                                 \\ \hline\hline
		\rowcolor{gray!30}                                                                                  & $0.067$                              & $0.066$                             & $0.135$                                    & $0.126$                                         & $0.066$                                            \\
		\rowcolor{gray!30}\multirow{-2}{1.5cm}{ \centering $f\nede$}                                        & $< 0.132$                            & $< 0.112$                           & $0.1330\pm 0.0255$                         & $0.1212^{+0.0275}_{-0.0244}$                    & $< 0.124$                                          \\ \hline
		\rowcolor{gray!30}                                                                                  & $2.543$                              & $2.496$                             & $2.583$                                    & $2.443$                                         & $2.496$                                            \\
		\rowcolor{gray!30}\multirow{-2}{1.5cm}{\centering\hspace*{-.3cm}$\log_{10}\left(\dfrac{m_\phi}{\textrm{Mpc}^{-1}}\right)$}                           & $2.554\pm 0.286$                     & $2.447^{+0.306}_{-0.168}$           & $2.553\pm 0.100$                           & $2.474\pm 0.113$                                & --                                                 \\ \hline
		\rowcolor{gray!30}                                                                                  & $349.486$                            & $313.191$                           & $382.974$                                  & $277.137$                                       & $313.191$                                          \\
		\rowcolor{gray!30}\rowcolor{gray!30}\multirow{-2}{1.5cm}{\centering $m_\phi$ {\tiny[Mpc${}^{-1}$]}} & $< 1090$                             & $< 745$                             & $367^{+70}_{-90}$                          & $308^{+50}_{-90}$                               & --                                                 \\ \hline
		\rowcolor{gray!30}                                                                                  & $4881.454$                           & $4593.696$                          & $5007.738$                                 & $4190.443$                                      & $4593.468$                                         \\
		\rowcolor{gray!30}\rowcolor{gray!30}\multirow{-2}{1.5cm}{\centering $z_*$}                          & $5270^{+1000}_{-2000}$               & $4640\pm 2000$                      & $4856^{+600}_{-600}$                       & $4417^{+500}_{-700}$                            & $4593.3^{+69.5}_{-52.1}$                           \\ \hline\hline
		\hspace*{.1cm}	Total $\chi^2$                                                                       & 3804.27                              & 4923.27                             & 3806.74                                    & 4926.58                                         & 4923.22                                            \\ \hline
		\hspace*{.1cm}	$\Delta\chi^2$                                                                       & -3.19                                & \cellcolor{violet!40} -3.32         & \cellcolor{teal!40}  -23.32                & \cellcolor{red!40} -23.13                       & \cellcolor{green!40} -3.37                         \\ \hline
		\hspace*{.1cm}	$H_0$ Tension                                                                        & $2.3\sigma$                          & $2.7\sigma$                         &                                   \multicolumn{2}{c|}{--}                                    & $2.4\sigma$                                        \\ \hline
		\hspace*{.1cm}	$Q_{\rm dmap}$                                                                       &                                      &                                     & $1.57\sigma$                               & $1.82\sigma$                                    &                                                    \\ \hline
	\end{tabular}

	\end{center}
	\caption{Table with the best-fit values followed by the posterior means and standard deviations of the $\Lambda$CDM and NEDE MCMC runs alternating datasets involving the baseline, SPT and S$H_0$ES. If only a one-sided constraint exists at $95\%$CL, it is given in the table instead of the $68\%$ error. This way of reporting, however, hides the fact that for fixed $m_\phi$ we do find almost a 2$\sigma$ evidence for $f\nede = 6.6^{+3.2}_{-3.9}\,\%$ even without S$H_0$ES, as further discussed in the text.}
	\label{tab:bestfitSPT}
\end{table}

Let us make some observations regarding issues faced when performing MCMC analyses using the SPT likelihood. The sampling issues for low fractions of $f\nede$ still persist when including SPT and excluding S$H_0$ES, this can be seen from the non-vanishing left tail in the marginal $f\nede$ distributions shown in Fig.~\ref{fig:triangleAltSPT} and from Fig.~\ref{fig:fNEDEvsH0SPT} where the contour of the run without fixing the trigger mass (purple) shows compatibility with vanishing $f\nede$ while the run with fixed mass (green) reaches higher values.  As discussed before, adapted sampling strategies are required to get rid of these spurious effects when sampling over all NEDE parameters. In any event, the crucial observation is that NEDE makes the datasets compatible with S$H_0$ES, corresponding to the green (and even purple) contour overlapping with the S$H_0$ES band within their respective $95\%$ confidence limits. This, in turn, justifies a joint analysis with S$H_0$ES. On a cautionary note, given the number of nuisance parameters related to the SPT likelihood, convergence is particularly slow for these runs.

Overall, we obtain consistent features with what we have seen from the baseline datasets and from ACT in the previous subsection.
It can be observed from the $\chi^2$ values reported in Tab.~\ref{tab:bestfitSPT}, that the NEDE model presents an improvement in $\chi^2$ for each combination of the datasets used. Most remarkable is the decrease of $\sim 23$ units (stronger than for ACT and similar to the baseline data) for the runs involving SPT and S$H_0$ES. As expected, NEDE manages to fit higher values of $H_0$ as can be seen from Fig.~\ref{fig:fNEDEvsH0SPT} which depicts the covariance contours for the different runs, including different combinations of SPT and S$H_0$ES. We, therefore, conclude that NEDE can also resolve the Hubble tension that arises between S$H_0$ES and the joint baseline (including Planck) and SPT data.

Including SPT has the effect of very slightly reducing the fraction of NEDE (by 1.2\%), yielding $f\nede = 12.1_{-2.4}^{+2.8}\, \%$ (as compared to $f\nede = 13.3 \pm 2.6 \%$). This still supports the existence of NEDE at $>99.7\%$. The marginal difference between both runs is also apparent from the broad overlap of the red with the blue contour in Fig.~\ref{fig:fNEDEvsH0SPT}, depicting the covariance between $f\nede$ and $H_0$. Similar insignificant changes occur for $\log_{10}(m_\phi)$ and $H_0$. As discussed before, the addition of NEDE is known to increase the amount of cold dark matter, $\omega_{\rm c}=\Omega_\mathrm{c} h^2$, and to give higher values for the scalar spectral index, $n_s$. These qualitative features are left unchanged by the addition of SPT, giving $\omega_\mathrm{c} = 0.129 \pm 0.003$ ($\Lambda$CDM: $\omega_\mathrm{c} = 0.117\pm 0.0008$) and $n_s = 0.989 \pm 0.006$  ($\Lambda$CDM: $n_s = 0.971 \pm 0.004$). For a more exhaustive comparison of the parameter contours against $\Lambda$CDM see Fig.~\ref{fig:triangleAltSPT} and for a comparison between SPT and ACT data within NEDE see Fig.~\ref{fig:compACTandSPT}. As can be seen, ACT and SPT have a similar and consistent impact on the $f\nede$ posteriors.

Without including S$H_0$ES but keeping SPT, we get an improvement of 3.3 units in $\chi^2$ when compared to a $\Lambda$CDM run with the same dataset combination and almost a 2$\sigma$ evidence for $f\nede = 6.6^{+3.2}_{-3.9}\,\%$. As opposed to the ACT runs discussed previously, including S$H_0$ES yields a stronger improvement of 23 units, which is similar to the baseline result.

As mentioned before, adding SPT does not affect the model's ability to address the Hubble tension, accordingly the  $Q_{\rm dmap}$ tension remains below $2\sigma$ (changing only insignificantly by $<0.3 \sigma$).

\begin{figure}[htbp]
	\centering
	\includegraphics[clip, width=.95\textwidth]{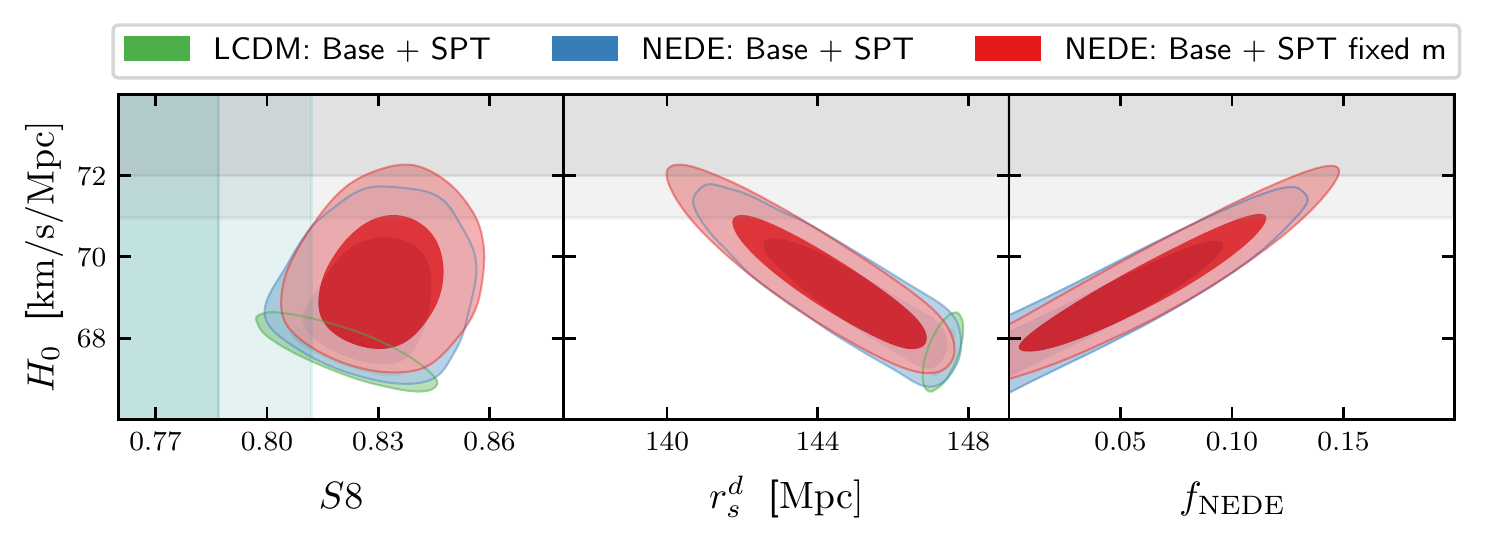}
	\label{fig:S8SPT}
	\caption{Plots showing the contours relating the Hubble parameter with $S8$, $r_s^{\rm d}$ and $f\nede$ as obtained when including SPT data while excluding the S$H_0$ES likelihood. The gray bands show the $1$ and $2\sigma$ contours coming from the latest S$H_0$ES reports \cite{Riess:2021jrx}, while the teal bands correspond to the measurements from KiDS+VIKING and DES \cite{Joudaki:2019pmv}.}
\end{figure}

As with ACT, we provide plots for $H_0$ against $S8$, $r_s^{\rm d}$ and $f\nede$ for the datasets that involve SPT but exclude S$H_0$ES in Fig.~\ref{fig:S8SPT}. We see that the NEDE contour overlaps separately with both the $S8$ and S$H_0$ES constraint at $95\%$ C.L.. However, we also see that there is only a very marginal common overlap between all three contours. This means that NEDE (in its current form) cannot solve both tensions simultaneously.

\subsection{Consequences for Inflation using BICEP}
\label{subsec:bicep}

In order to explain the homogeneity, isotropy, and flatness of the observed Universe, an epoch of cosmological inflation has been added to the $\Lambda$CDM model as suggested in the early 80's \cite{Starobinsky:1980te, Guth:1980zm, Linde:1981mu}. Since then, many variants and models specifying the details of how this epoch transpired have been proposed. They can be broadly classified into two categories according to the number of additional fields required in a given inflationary model. In the simplest cases, i.e.~ for any model that implements inflation by means of a single additional field, it is possible to compute the predicted primordial power spectrum and constrain $n_s$, the scalar spectral tilt, and $r$, the tensor-to-scalar ratio.

With ongoing experiments such as Planck \cite{Planck:2018jri} and the BICEP3/Keck Array \cite{BICEP:2021xfz}, $n_s$ and $r$ can be constrained once a cosmological model is assumed. While Planck measures the whole sky, BICEP3 focuses on smaller low-foreground patches of the sky and achieves higher sensitivity. Most inflationary models can be constrained by the parameters $n_s$ and $r$, this is, in particular, true for single-field models observing the slow-roll single-field condition. At present, Starobinsky inflation \cite{Starobinsky:1980te}, or $R^2$ inflation, predicts values that are in very good agreement with $\Lambda$CDM, while many other models for inflation are believed to be excluded given the amount of confidence in the standard model of cosmology -- $\Lambda$CDM.

\begin{figure}[htbp]
	\centering
	\includegraphics[clip, width=0.9\textwidth, trim=0 2mm 0 0]{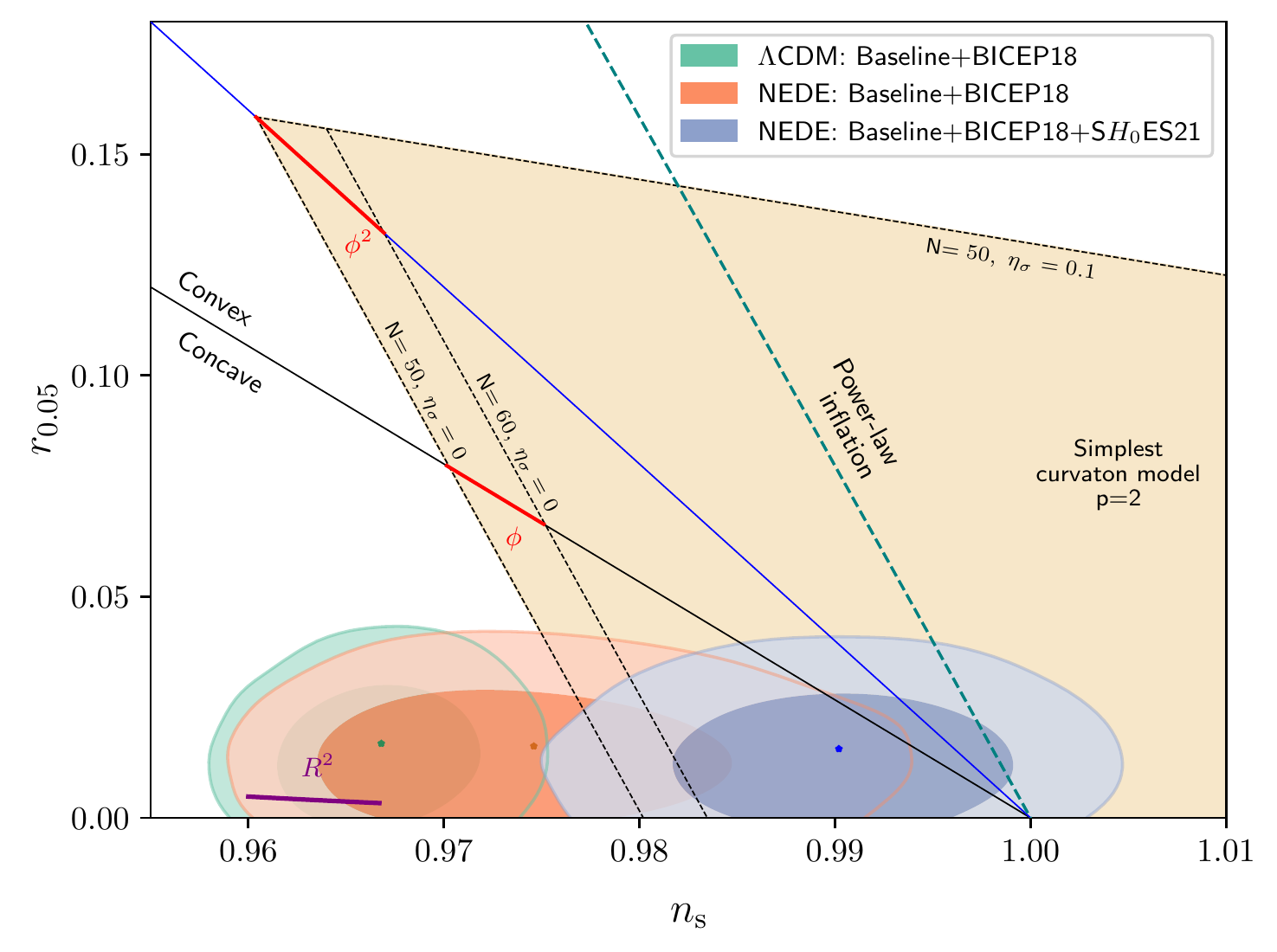}
	\caption{Results for the $68 \%$ and $95\%$ C.L. contours relating $n_s$ and $r$ at a pivot scale of $k_*=0.05$, for the $\Lambda$CDM and NEDE models alternating the baseline datasets with S$H_0$ES while including BICEP18. The small asterisks represent the mean posterior value of the corresponding contours.}
	\label{fig:n_s-rPlot}
\end{figure}

For the reasons above, it is extremely important to update these constraints in view of the NEDE model. To that end, we perform an MCMC simulation to obtain the NEDE confidence contours in the $n_s-r$ plane.
We employ the baseline datasets supplemented with S$H_0$ES 2021 \cite{Riess:2021jrx} and BICEP18 \cite{BICEP:2021xfz} and evaluate primordial quantities at the pivot scale $k_* = 0.05$ Mpc${}^{-1}$. We sample $r$ while keeping the single-field consistency relation $n_t = - r/8$. We demand the sampling to reach a Gelman-Rubin criterion of  $R-1<0.05$. Our results are summarized in Fig.~\ref{fig:n_s-rPlot}, which shows the $68 \%$  (dark shading) and $95\%$ (light shading) C.L. of one $\Lambda$CDM  (green) and two NEDE simulations (orange and blue) along with the predicted values and regions for some relevant models of inflation.
Best-fit values, posterior means, and errors can be found in the Tab.~\ref{tab:bicepBestfit}.

Looking at Fig.~\ref{fig:n_s-rPlot}, we observe that the effect of assuming NEDE is to shift the contours towards a bluer spectrum (or larger $n_s$ equivalently). In particular, when considering the baseline dataset together with BICEP18 and S$H_0$ES, values of $n_s$ very close to unity can be reached within $68 \%$ C.L.. While the current bounds on $r$ exclude the simplest $\phi$ and $\phi^2$ models of inflation, NEDE allows $n_s$ to be higher and rescues models that have convex potentials, power-law inflation, or that alternatively, feature more than one field.

Besides the simpler $\phi^2$ potential line (blue solid) and the convex line  (black solid), we see that power-law inflation \cite{Lucchin:1984yf, Liddle:1988tb}  (green dashed) is again possible within the NEDE model. In such a case, the scale factor of the universe evolves following a power law and is implemented via a potential that is exponential, and in agreement with a nearly conformal initial state of the Universe \cite{DAmico:2022agc}.

It is worth looking at two-field models as well, in view of the NEDE posteriors presented in Fig.~\ref{fig:n_s-rPlot}. An estimate of the contours shown has already appeared in the context of a two-stage monodromy inflation model \cite{DAmico:2021fhz}, where these seem to be compatible with the first stage of inflation lasting about 55 to 60 $e$-folds. Along similar lines, multi-field models featuring an inflation together with a curvaton as proposed in \cite{Enqvist:2001zp, Lyth:2001nq, Moroi:2001ct}, can cover a bigger piece of the parameter space for $n_s$ and $r$,  corresponding to the yellow shaded area in Fig.~\ref{fig:n_s-rPlot}. Following the constraints in the general mixed inflaton-curvaton ~\cite{Langlois:2004nn,Ferrer:2004nv,Easson:2010uw,Kinney:2012ik,Fujita:2014iaa}, we can estimate the number of $e$-folds and the $\eta_\sigma$ parameter proportional to the curvaton's mass for the mean values produced by NEDE. For the case of the simplest curvaton model \cite{Bartolo:2002vf} one has that the scalar spectral tilt and the tensor-to-scalar ratio are given by
\begin{subequations}
\begin{align}
	n_{\rm s} &= 1 -\frac{1}{1+R}\frac{2(2+p)}{4 N + p} + \frac{R}{1+R}\bigg[ -2\epsilon + 2\eta_\sigma\bigg] \,,\\
	r &= \frac{16\epsilon}{1+R}\,,
\end{align}
\end{subequations}
%where a chaotic inflation potential $V(\phi) = \frac{1}{p}\frac{\phi^p}{M^{p-4}}$ was assumed, $\epsilon = \frac{M_{\rm Pl}^2}{2}\frac{V'^2}{V^2}$ is a slow-roll parameter, $R$ is the ratio of the curvaton-to-inflaton power spectra, and $N$ is the number of e-folds. The yellow shaded area in Figure~\ref{fig:n_s-rPlot} is produced by varying the number of $e$-folds, $N\in[50,60]$, and the mass parameter $\eta_\sigma\in[0,0.1]$. As can be seen from the figure, keeping $\eta_\sigma$ fixed and increasing the number of $e$-folds displaces the values to the right, while increasing $\eta_\sigma$ decreases the slope of a fixed $N$ line (effectively sweeping the yellow region). In this way, each combiacanation of $n_s$ and $r$ in the simplest curvaton model is associated with a range of $N$ and $\eta_\sigma$ values. We can estimate using the means for the NEDE simulation with all the datasets, $n_s=0.9902$ and $r=0.0154$, and for inflation lasting between 50 to 60 e-folds, that the mass parameter would be between 0.005 and 0.007 which puts the mass of the curvaton with $\eta_\sigma \sim 0.007$ for $N=50$ and $\eta_\sigma \sim 0.005$ for $N=60$ at around
\begin{equation}
	m_\sigma = H_* \sqrt{3\eta_\sigma} \sim 10^{13} - 10^{14} \;\text{GeV},
\end{equation}
depending on the number of $e$-folds chosen. Having an extra light scalar field, like the curvaton, with a mass within an order of magnitude of the Hubble scale during inflation does not add any additional fine-tuning or naturalness problems.

It is interesting to note that in the case of $r=0$, the pure curvaton model predicts a testable level of local non-Gaussianity of $f_{\textnormal{NL}} = -5/4$ \cite{Bartolo:2003jx}, while for larger $r$, $f_{\textnormal{NL}}$ becomes proportionally smaller. Hence the tensor-to-scalar ratio and non-Gaussianity are complementary probes of the curvaton model.

\begin{table}[htbp]
	\begin{center}
	\renewcommand{\arraystretch}{1.45}
	\setlength\tabcolsep{0pt}
	\fontsize{8}{11}\selectfont
	\begin{tabular}{|>{\centering\arraybackslash}p{3.1cm}|>{\centering\arraybackslash}p{2.5cm}|>{\centering\arraybackslash}p{2.5cm}|>{\centering\arraybackslash}p{2.5cm}|}
		\hline
		\multirow{2}{3cm}{\hspace*{.2cm}{\bf Parameter Name}}              & \multirow{2}{2cm}{$\Lambda$CDM: Base + BICEP18} & \multirow{2}{1.7cm}{NEDE: Base + BICEP18} & \multirow{2}{1.7cm}{NEDE: Base + BICEP18 + S$H_0$ES} \\
		                                                                   &                                                 &                                           &                                                      \\ \hline\hline
		\multirow{2}{1.5cm}{\centering $\Omega_b h^2$}                     & $0.022$                                         & $0.023$                                   & $0.023$                                              \\
		                                                                   & $0.0224\pm 0.00013$                             & $0.0226^{+0.0002}_{-0.0003}$              & $0.0230\pm 0.0002$                               \\ \hline
		\multirow{2}{1.5cm}{\centering $\Omega_{\rm c} h^2$}               & $0.119$                                         & $0.124$                                   & $0.130$                                              \\
		                                                                   & $0.1193\pm 0.0009$                              & $0.1233^{+0.0019}_{-0.0042}$              & $0.1299\pm 0.0032$                                 \\ \hline
		\multirow{2}{1.5cm}{\centering $H_0$}                              & $67.680$                                        & $69.078$                                  & $71.374$                                             \\
		                                                                   & $67.688\pm 0.408$                               & $68.963^{+0.720}_{-1.35}$                 & $71.538\pm 0.864$                                    \\ \hline
		\multirow{2}{1.5cm}{\centering $\log(10^{10} A_{\rm s})$}          & $3.046$                                         & $3.063$                                   & $3.062$                                              \\
		                                                                   & $3.0487\pm 0.0142$                              & $3.0550\pm 0.0152$                        & $3.0685\pm 0.0147$                                   \\ \hline
		\multirow{2}{1.5cm}{\centering $n_s$}                              & $0.965$                                         & $0.975$                                   & $0.988$                                              \\
		                                                                   & $0.9668\pm 0.0037$                              & $0.9746^{+0.0060}_{-0.0089}$              & $0.9902\pm 0.0062$                                 \\ \hline
		\multirow{2}{1.5cm}{\centering $\tau_{\rm reio}$}                  & $0.054$                                         & $0.057$                                   & $0.054$                                              \\
		                                                                   & $0.0569\pm 0.0072$                              & $0.0565\pm 0.0073$                        & $0.0578^{+0.0069}_{-0.0076}$                      \\ \hline
		\multirow{2}{1.5cm}{\centering $r_{0.05}$}                         & $0.000$                                         & $0.000$                                   & $0.000$                                              \\
		                                                                   & $< 0.0354$                                      & $< 0.0344$                                & $< 0.0335$                                           \\ \hline
		\rowcolor{gray!30}                                                 & --                                              & $0.054$                                   & $0.126$                                              \\
		\rowcolor{gray!30}\multirow{-2}{1.5cm}{\centering $f\nede$ }       & --                                              & $< 0.118$                                 & $0.129^{+0.030}_{-0.026}$                         \\ \hline
		\rowcolor{gray!30}                                                 & --                                              & $2.568$                                   & $2.527$                                              \\
		\rowcolor{gray!30}\multirow{-2}{1.5cm}{\centering \hspace*{-.3cm}$\log_{10}\left(\dfrac{m_\phi}{\textrm{Mpc}^{-1}}\right)$} & --                                              & $2.417^{+0.389}_{-0.173}$                 & $2.511^{+0.143}_{-0.097}$                           \\ \hline
		\rowcolor{gray!30}                                                 & --                                              & $2.012$                                   & $2.090$                                              \\
		\rowcolor{gray!30}\multirow{-2}{1.5cm}{\centering $3\omega\nede$ } & --                                              & $> 1.42$                                  & $2.119^{+0.120}_{-0.195}$                            \\ \hline
		\rowcolor{gray!30}                                                 & --                                              & $370.021$                                 & $336.556$                                            \\
		\rowcolor{gray!30}\multirow{-2}{1.5cm}{\centering $m_\phi$[Mpc${}^{-1}$]}       & --                                              & $< 781$                                   & $339^{+90}_{-100}$                                   \\ \hline
		\rowcolor{gray!30}                                                 & --                                              & $5065.098$                                & $4673.963$                                           \\
		\rowcolor{gray!30}\multirow{-2}{1.5cm}{\centering $z_*$ }          & --                                              & $4569\pm 2000$                            & $4633\pm 800$                                        \\ \hline
	\end{tabular}
\end{center}
\caption{Best-fit values, means and $1\sigma$ confidence intervals for MCMC runs involving the BICEP18 dataset, together with their corresponding $\chi^2$ individual and total values.}
\label{tab:bicepBestfit}
\end{table}

\section{Discussion and summary}

\begin{figure}[htbp]
	\centering
	\includegraphics[clip, width=.99\textwidth]{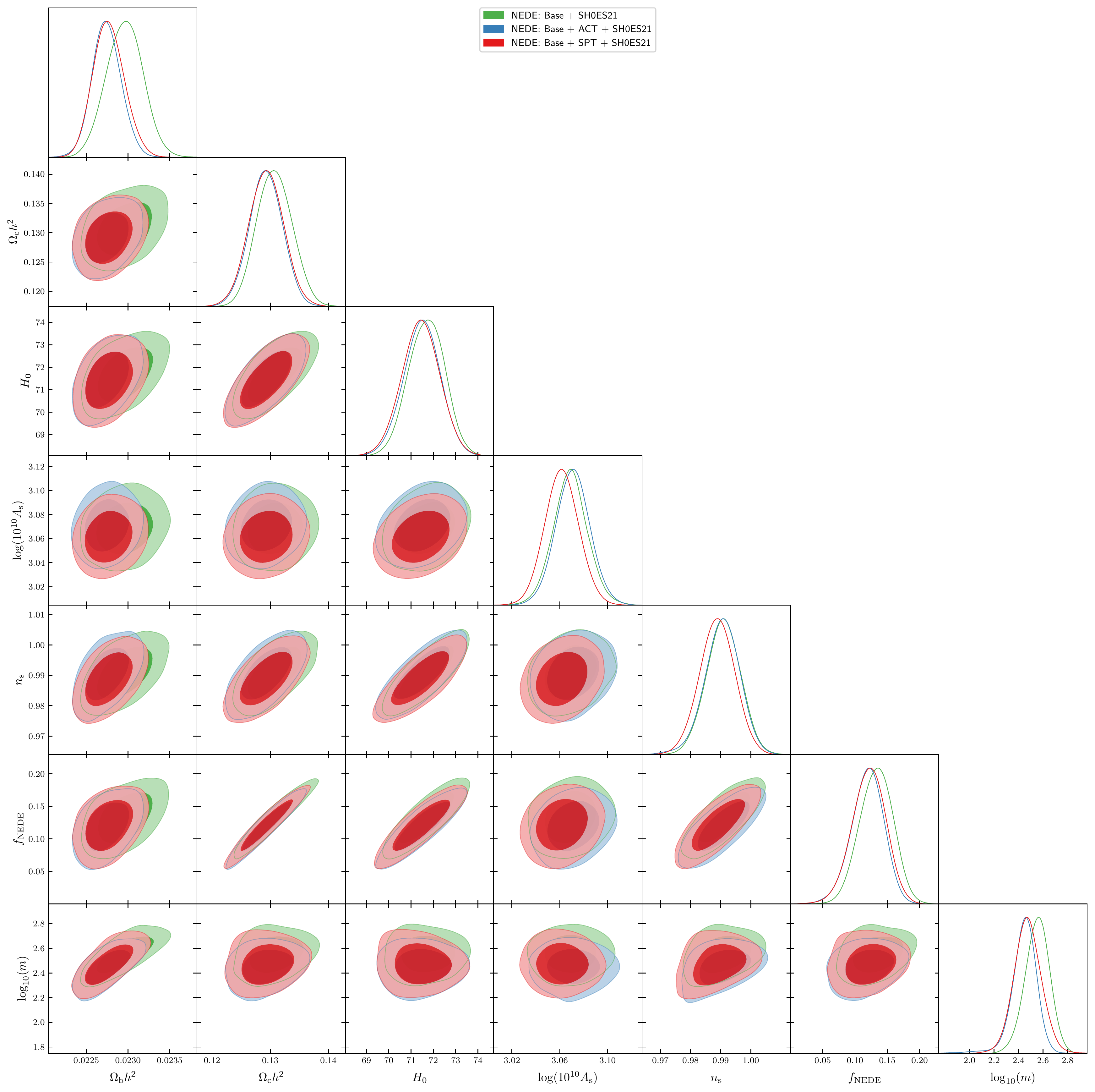}
	\caption{Contour plots for some of the parameters sampled in the NEDE model. The two sets of contours show the impact of SPT and ACT on the predicted cosmological parameter posteriors for the NEDE model. Overall, adding these additional datasets leads to only small changes.}
	\label{fig:compACTandSPT}
\end{figure}

We tested the NEDE model against additional CMB data arising from the ground-based experiments ACT, SPT, and BICEP3/Keck. We have observed in all three cases that the NEDE model maintains its ability to address the Hubble tension despite the additional constraining power of these datasets.  In particular, we found that each ground-based experiment in combination with our baseline dataset, consisting of BAO, Pantheon and Planck data, prefers a non-vanishing fraction of NEDE with a significance larger than $4.5 \sigma $ if we include a S$H_0$ES prior on $H_0$.  In all cases, this comes with a significantly reduced Hubble tension and a more than $ 4 \sigma$ increase of the spectral tilt  $n_s$ as compared to its $\Lambda$CDM value. Overall, we find that the extracted parameter values are fully compatible with the ones obtained by only using CMB data from Planck. This is emphatically summarized in Fig.~\ref{fig:compACTandSPT}.

In a first series of simulations, we have examined the effects of including the CMB likelihood from ACT. We first recovered a slight bi-modality in the three-parameter NEDE model that had previously been reported in~\cite{Poulin:2021bjr}. This is in general a challenge for sampling algorithms and we dealt with it by choosing the dominant mode centered around an equation of state parameter $w_\mathrm{NEDE} =2/3$. This value falls within the theoretically allowed regime and corresponds to a bubble wall condensate that decays quicker than the dominant radiation component. In the presence of ACT, this two-parameter model  was found to make the inferred value of $H_0$ compatible with the S$H_0$ES value, motivating a combined analysis with S$H_0$ES. In this case, we report $71.49\pm 0.822$ km/s/Mpc ($68\%$ C.L.) and a best-fit of $H_0 = 72.09$ km/s/Mpc. The fit quality compared to $\Lambda$CDM improves significantly by $\Delta \chi^2 = -15.9$, albeit less pronounced than without ACT ($\Delta \chi^2 = -23.3$). This corresponds to a reduced $Q_{\rm dmap}$ tension of  $2.9 \sigma$ down from $4.8 \sigma$ within $\Lambda$CDM, which should be compared with a tension of $1.6 \sigma$ for the baseline dataset. We therefore conclude that ACT data makes NEDE somewhat less efficient at addressing the Hubble tension, although the evidence for NEDE remains with $4.7 \sigma$ high. At this stage, it remains to be seen whether this observation is supported by future data releases.

With SPT things are even simpler. Both the posteriors and the $\chi^2$ improvement over $\Lambda$CDM only change slightly compared to the baseline runs. For completeness, we find $71.431\pm 0.852$ km/s/Mpc ($68\%$ C.L.) and a best-fit of $H_0=71.77$ km/s/Mpc along with a strong fit corresponding to $\Delta \chi^2 = -23.1$. Moreover, the $Q_{\rm dmap}$ remains below $2 \sigma$. We stress that this outcome, while it might appear uneventful, is important news for NEDE. After all, CMB data is the best probe we have to test the model. It also puts the marginally weaker ACT outcome into perspective and highlights the relevance of testing the model with different probes.

Our analysis  also shows that both ACT and SPT data do not yet have the constraining power to strengthen (or weaken) the evidence for NEDE. We expect this conclusion to change with future data. Also the situation with $S8$ remains unchanged. NEDE in its current form does not improve on the issue, while also not making it much worse compared to $\Lambda$CDM. Here, progress might come from two directions: First, with Hot NEDE~\cite{Niedermann:2021vgd,Niedermann:2021ijp}  there is a new microscopic implementation of NEDE that contains a richer dark sector that is also expected to change the prediction for $S8$. Alternatively, given that the tension is mild, there is still the possibility of systematics explaining the data. For example, there are hints from the James Webb Space Telescope indicating anomalies within $\Lambda$CDM, which themselves may point towards a higher $S8$ \cite{Boylan-Kolchin:2022kae}. Moreover, possible systematic issues in the lensing and clustering probes are currently being studied  \cite{Amon:2022ycy} and may still have something to say about $S8$.

Finally, we also explored the relevance of our findings for inflation. Specifically, we supplemented our analysis with BICEP3/Keck data to constrain the tensor-to-scalar ratio and create an updated version of the $r$ vs $n_s$ plot, which is the main phenomenological input for inflationary model building. In agreement with a previous statement in the literature~\cite{Niedermann:2020dwg,DAmico:2021fhz}, we find that NEDE prefers a significantly bluer spectrum, $n_s = 0.9902\pm 0.006$. If this result stands the test of time, it will have dramatic consequences for our understanding of the primordial universe. Specifically, it implies that Starobinsky's model is no longer a good fit to the data. Instead, the simplest curvaton model would provide a better explanation for the initial conditions of the Universe in agreement with NEDE as a solution to the Hubble tension.

\subsection*{Acknowledgements}
We would like to thank Guido D'Amico, Nemanja Kaloper, Antony Lewis, Vivian Poulin, Toni Riotto, and Jussi Valiviita for useful discussions. This work is supported by  Independent Research Fund Denmark grant 0135-00378B.

\appendix

\section{Complementary results for MCMC analyses of the different runs}
\label{app:posteriorsDeviations}

\begin{figure}[htbp]
	\centering
	\includegraphics[clip, width=\textwidth]{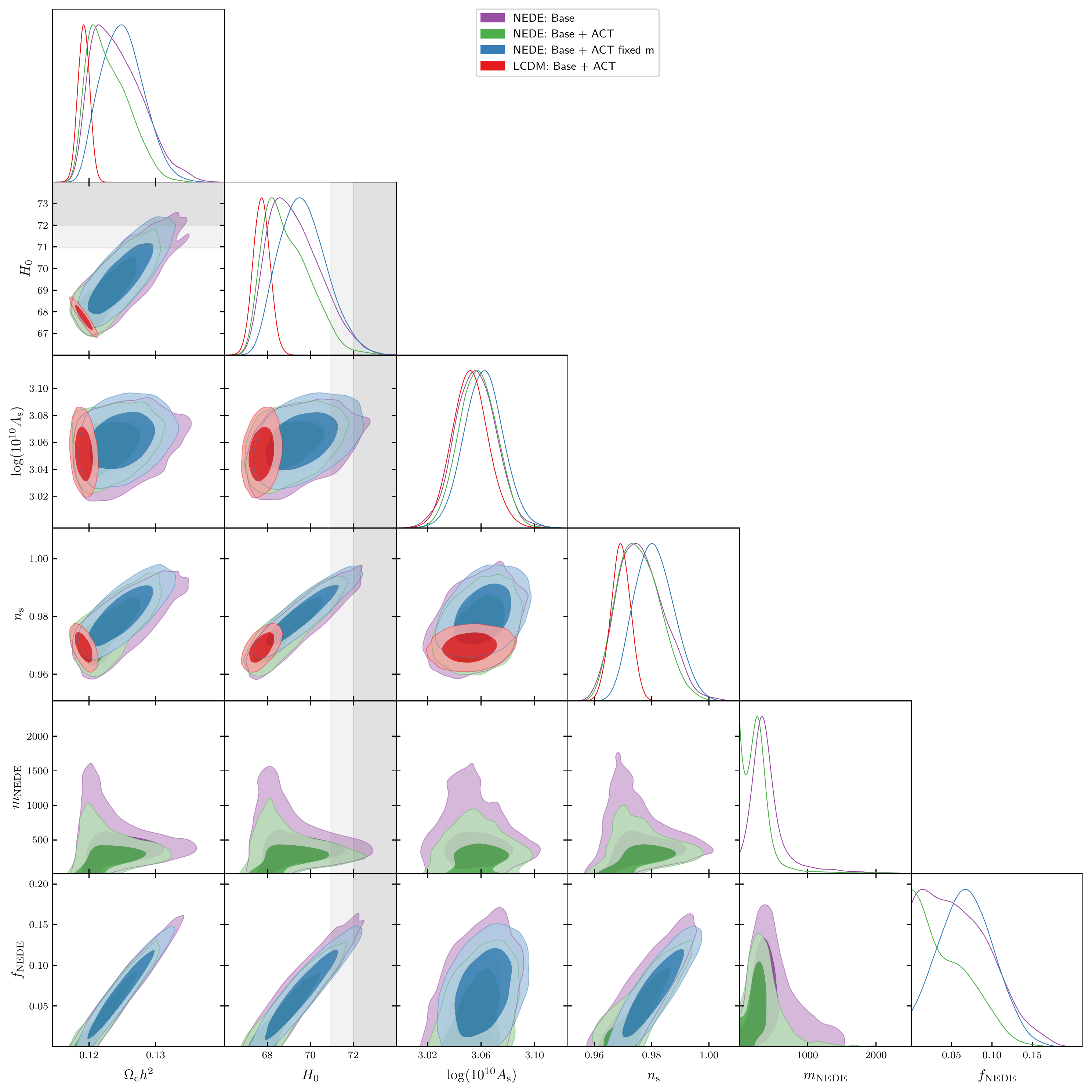}
	\caption{Triangle plot displaying the original six parameters of $\Lambda$CDM and their $68 \%$ and $95\%$ C.L. contours coming out of the MCMC analysis of $\Lambda$CDM and the NEDE models, highlighting the impact of the ACT dataset on the fits, described in subsection \ref{subsec:nedeACT}.}
	\label{fig:triangleAltACT}
\end{figure}

\begin{figure}[htbp]
	\centering
	\includegraphics[clip, width=\textwidth]{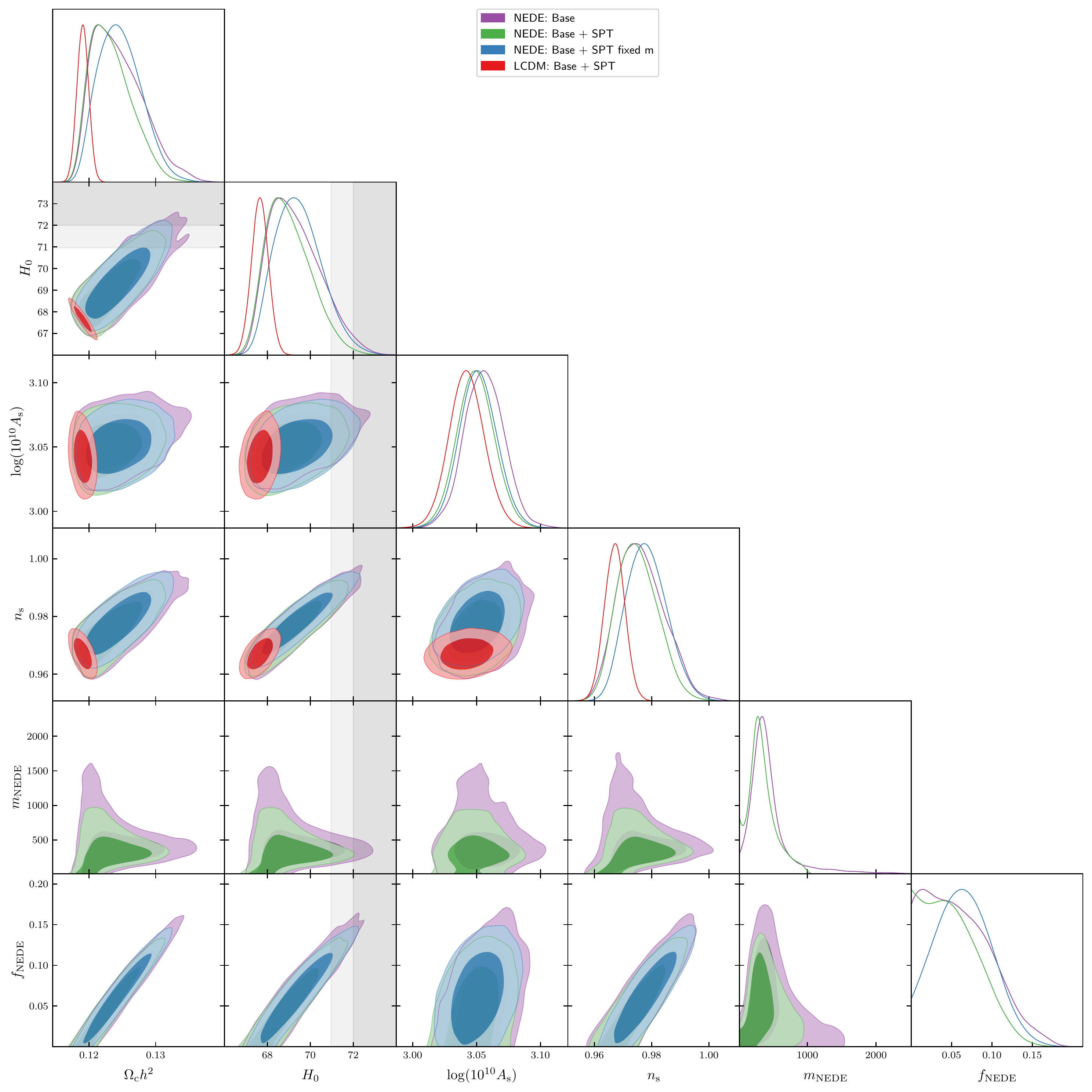}
	\caption{Triangle plot displaying the original six parameters of $\Lambda$CDM and their $68 \%$ and $95\%$ C.L. contours coming out of the MCMC analysis of $\Lambda$CDM and the NEDE models, highlighting the impact of the SPT dataset on the fits. This set of simulations is described in subsection \ref{subsec:nedeSPT}.}
	\label{fig:triangleAltSPT}
\end{figure}

\begin{table}[htbp]
	\centering
	\renewcommand{\arraystretch}{1.5}
	\setlength\tabcolsep{0pt}
	\fontsize{8}{11}\selectfont
	\begin{tabular}{|c|c|c|}
		\hline
		\hspace*{.5cm} Parameter Name \hspace*{.5cm} & \hspace*{.5cm}  Mean and Std. Dev.   \hspace*{.5cm} & \hspace*{.5cm} Best-fit \hspace*{.5cm} \\ \hline\hline
		            $\Omega_\mathrm{m}$              &                $0.2991\pm 0.0051$                 &                $0.299$                 \\
		          $\Omega_\mathrm{m} h^2$            &                $0.1517\pm 0.0031$                 &                $0.1529$                \\
		              $\Omega_\Lambda$               &                $0.7008\pm 0.0051$                 &                $0.7009$                \\
		                 $\sigma_8$                  &                $0.8386\pm 0.0096$                 &                $0.8461$                \\
		             $\sigma_8/h^{0.5}$              &                $0.9936\pm 0.0087$                 &                $1.0007$                \\
		     $\sigma_8 \Omega_\mathrm{m}^{0.5}$      &                $0.4586\pm 0.0065$                 &               $0.4627 $                \\
		    $\sigma_8 \Omega_\mathrm{m}^{0.25}$      &                $0.6201\pm 0.0075$                 &               $0.6257 $                \\
		         ${\rm{Age}}/\mathrm{Gyr}$           &             $13.24^{+0.13}_{-0.15}$              &                $ 13.19$                \\
		              $r_s^\mathrm{d}$               &              $142.0^{+1.37}_{-1.57}$               &                $141.46$                \\ \hline
	\end{tabular}
	\caption{Posterior mean and standard deviation for some derived parameters of the run described in Sec.~\ref{subsec:freeEOS} having an unconstrained equation of state.}
	\label{tab:derivedFreeEOS}
\end{table}

Simulations were all made with fixed EOS and for the following combinations of likelihoods:
\begin{itemize}
	\item R1 - NEDE: Baseline
	\item R2 - NEDE: Baseline + ACT
	\item R3 - NEDE: Baseline + S$H_0$ES21
	\item R4 - NEDE: Baseline + ACT + S$H_0$ES21
	\item R5 - NEDE: Baseline + ACT fixed m to best-fit of R2
\end{itemize}
\begin{table}[htbp]
	\centering
	\renewcommand{\arraystretch}{1.5}
	\setlength\tabcolsep{0pt}
	\fontsize{8}{11}\selectfont
	\begin{tabular}{c|c|c|c|c|c}
		               Name                 & \hspace{.1cm} R1 - Mean$_{\rm Lower}^{\rm Upper}$ \hspace{.1cm} & \hspace{.1cm} R2 - Mean$_{\rm Lower}^{\rm Upper}$ \hspace{.1cm} & \hspace*{.1cm} R3 - Mean$_{\rm Lower}^{\rm Upper}$ \hspace*{.1cm} & \hspace{.1cm}  R4 - Mean$_{\rm Lower}^{\rm Upper}$ \hspace{.1cm} & \hspace{.1cm}R5 - Mean$_{\rm Lower}^{\rm Upper}$ \hspace{.1cm} \\ \hline\hline
		        $\Omega_\mathrm{m}$         &                       $0.3075\pm 0.0061$                        &                       $0.3074\pm 0.0059$                        &                        $0.3003\pm 0.0051$                         &                        $0.2987\pm 0.0050$                        &                       $0.3055\pm 0.0057$                       \\
		      $\Omega_\mathrm{m} h^2$       &                  $0.1477^{+0.0026}_{-0.0050}$                   &                  $0.1461^{+0.00212}_{-0.0041}$                  &                        $0.1544\pm 0.0031$                         &                        $0.1527\pm 0.0029$                        &                  $0.1483^{+0.0029}_{-0.0037}$                  \\
		         $\Omega_\Lambda$           &                       $0.6924\pm 0.0061$                        &                       $0.6925\pm 0.0059$                        &                        $0.6996\pm 0.0051$                         &                        $0.7012\pm 0.0050$                        &                       $0.6944\pm 0.0057$                       \\
		          $z_\mathrm{re}$           &                        $7.905\pm 0.726$                         &                        $7.744\pm 0.703$                         &                         $8.068\pm 0.734$                          &                         $7.927\pm 0.738$                         &                        $7.769\pm 0.725$                        \\
		            $\sigma_8$              &                   $0.8237^{+0.0095}_{-0.013}$                   &                   $0.8225^{+0.0084}_{-0.012}$                   &                        $0.8415\pm 0.0092$                         &                        $0.8408\pm 0.0092$                        &                  $0.8291^{+0.0096}_{-0.011}$                   \\
		        $\sigma_8/h^{0.5}$          &                       $0.9895\pm 0.0091$                        &                       $0.9906\pm 0.0084$                        &                        $0.9938\pm 0.0088$                         &                        $0.9945\pm 0.0087$                        &                       $0.9933\pm 0.0084$                       \\
		$\sigma_8 \Omega_\mathrm{m}^{0.5}$  &                       $0.4568\pm 0.0070$                        &                       $0.4560\pm 0.0062$                        &                        $0.4611\pm 0.0067$                         &                        $0.4595\pm 0.0064$                        &                       $0.4583\pm 0.0061$                       \\
		$\sigma_8 \Omega_\mathrm{m}^{0.25}$ &                  $0.6134^{+0.0076}_{-0.0093}$                   &                  $0.6124^{+0.0068}_{-0.0080}$                   &                        $0.6229\pm 0.0075$                         &                        $0.6216\pm 0.0073$                        &                       $0.6164\pm 0.0073$                       \\
		     ${\rm{Age}}/\mathrm{Gyr}$      &                    $13.5^{+0.255}_{-0.117}$                     &                    $13.6^{+0.216}_{-0.090}$                     &                          $13.1\pm 0.127$                          &                     $13.2^{+0.118}_{-0.131}$                     &                    $13.5^{+0.188}_{-0.148}$                    \\
		         $r_s^\mathrm{d}$           &                     $144.3^{+2.65}_{-1.23}$                     &                     $145.2^{+2.20}_{-0.98}$                     &                         $140.73\pm 1.40$                          &                         $141.54\pm 1.36$                         &                    $144.0^{+1.88}_{-1.53}$                     \\ \hline
	\end{tabular}
	\label{tab:derivedPosteriorsACT}
	\caption{Posterior means for some derived parameters in the NEDE MCMC simulations for different combinations of datasets involving the baseline, ACT and S$H_0$ES, with a fixed EOS to $\omega\nede=2/3$.}
\end{table}

Posterior mean values and the 68.27\% confidence intervals for the set of runs involving the SPT data
\begin{itemize}
	\item S1 - $\Lambda$CDM: Planck BAO SN SPT
	\item S2 - NEDE: Baseline + SPT fixed EOS
	\item S3 - NEDE: Baseline + S$H_0$ES
	\item S4 - NEDE: Baseline + SPT fixed EOS + S$H_0$ES
	\item S5 - NEDE: Baseline + SPT fixed EOS fixed $m_\phi$ to best-fit of S2
\end{itemize}

\begin{table}[htbp]
	\centering
	\renewcommand{\arraystretch}{1.5}
	\setlength\tabcolsep{0pt}
	\fontsize{8}{11}\selectfont
	\begin{tabular}{c|c|c|c|c|c}
				Name                 & \hspace{.1cm} S1 - Mean$_{\rm Lower}^{\rm Upper}$ \hspace{.1cm} & \hspace{.1cm} S2 - Mean$_{\rm Lower}^{\rm Upper}$ \hspace{.1cm} & \hspace{.1cm} S3 - Mean$_{\rm Lower}^{\rm Upper}$ \hspace{.1cm} & \hspace{.1cm}S4 - Mean$_{\rm Lower}^{\rm Upper}$\hspace{.1cm} & \hspace{.1cm} S5 - Mean$_{\rm Lower}^{\rm Upper}$ \hspace{.1cm} \\ \hline\hline
		        $\Omega_\mathrm{m}$         & $0.3105\pm 0.0053$  &      $0.3077\pm 0.0059$      & $0.3003\pm 0.0051$ & $0.2994\pm 0.0052$ &      $0.3066\pm 0.0057$       \\
		      $\Omega_\mathrm{m} h^2$       & $0.1421\pm 0.0009$ & $0.1465^{+0.0021}_{-0.0041}$ & $0.1544\pm 0.0031$ & $0.1527\pm 0.0031$ & $0.1479^{+0.0028}_{-0.0039}$ \\
		         $\Omega_\Lambda$           & $0.6894\pm 0.0053$  &      $0.6923\pm 0.0059$      & $0.6996\pm 0.0051$ & $0.7006\pm 0.0052$ &      $0.6933\pm 0.0057$       \\
		          $z_\mathrm{re}$           &  $7.687\pm 0.706$   &       $7.708\pm 0.723$       &  $8.068\pm 0.73$   &  $7.885\pm 0.721$  &        $7.697\pm 0.72$        \\
		            $\sigma_8$              & $0.8075\pm 0.0057$  & $0.8188^{+0.0084}_{-0.011}$  & $0.8415\pm 0.0092$ & $0.8357\pm 0.0092$ & $0.8227^{+0.0090}_{-0.011}$ \\
		        $\sigma_8/h^{0.5}$          & $0.9818\pm 0.0079$  &      $0.9856\pm 0.0084$      & $0.9938\pm 0.0088$ & $0.9888\pm 0.0085$ &      $0.9871\pm 0.0084$      \\
		$\sigma_8 \Omega_\mathrm{m}^{0.5}$  & $0.4500\pm 0.0054$  &      $0.4541\pm 0.0062$      & $0.4611\pm 0.0067$ & $0.4572\pm 0.0064$ &      $0.4555\pm 0.0061$      \\
		$\sigma_8 \Omega_\mathrm{m}^{0.25}$ & $0.6028\pm 0.0054$  & $0.6098^{+0.0068}_{-0.0077}$ & $0.6229\pm 0.0075$ & $0.6181\pm 0.0073$ &      $0.6122\pm 0.0073$      \\
		     ${\rm{Age}}/\mathrm{Gyr}$      &  $13.80\pm 0.018$   &   $13.56^{+0.22}_{-0.094}$   &  $13.14\pm 0.13$   &  $13.20\pm 0.132$  &   $13.49^{+0.197}_{-0.143}$   \\
		         $r_s^\mathrm{d}$           &   $147.3\pm 0.22$   &   $144.9^{+2.25}_{-1.00}$    &  $140.7\pm 1.40$   &  $141.5\pm 1.42$   &    $144.2^{+1.98}_{-1.48}$    \\ \hline
	\end{tabular}
	\label{tab:derivedPosteriorsSPT}
	\caption{Table displaying the posterior means and standard deviations for some derived parameters of the runs involving the SPT likelihood.}
\end{table}

\begin{table}[htbp]
	\begin{center}
		\renewcommand{\arraystretch}{1.14}
		\setlength\tabcolsep{0pt}
		\fontsize{8}{11}\selectfont

		\begin{tabular}{|c||c|c|c|c|c|c|}
			\hline
			\multirow{3}{1.5cm}{Parameter Name} &                                                                                                          \multicolumn{6}{c|}{$\Lambda$CDM (Base = Planck+BAO+SN)}                                                                                                           \\
			        \hhline{~|------|}          & \multirow{2}{1.5cm}{\centering Base} &   \multirow{2}{1.5cm}{\centering+ACT}   & \multirow{2}{1.5cm}{\centering+SPT} & \multirow{2}{1.5cm}{\centering+S$H_0$ES} & \multirow{2}{1.5cm}{\centering+ACT +S$H_0$ES21} & \multirow{2}{1.5cm}{\centering+SPT +S$H_0$ES21} \\
			                                    &                                      &                                         &                                     &                                          &                                                 &                                                 \\ \hline\hline
			      $\Omega_\mathrm{b} h^2$       &                0.022                 &                  0.022                  &                0.022                &                  0.023                   &                      0.022                      &                      0.023                      \\
			      $\Omega_\mathrm{c} h^2$       &                0.119                 &                  0.119                  &                0.119                &                  0.117                   &                      0.118                      &                      0.118                      \\
			               $H_0$                &                67.643                &                 67.817                  &               67.747                &                  68.630                  &                     68.242                      &                     68.361                      \\
			      $\;\log(10^{10} A_s)\;$       &                3.052                 &                  3.067                  &                3.038                &                  3.061                   &                      3.073                      &                      3.059                      \\
			               $n_s$                &                0.966                 &                  0.968                  &                0.967                &                  0.971                   &                      0.971                      &                      0.972                      \\
			         $\tau_{\rm reio}$          &                0.058                 &                  0.057                  &                0.052                &                  0.062                   &                      0.060                      &                      0.061                      \\ \hline
			   \hspace*{.1cm}	Total $\chi^2$    &               3807.45                & \cellcolor{violet!40}           4042.38 &   \cellcolor{green!40}   4926.59    &    \cellcolor{teal!40}        3830.06    &      \cellcolor{red!40}           4064.94       &    \cellcolor{red!40}               4949.71     \\ \hline
			   \hspace*{.1cm}	$Q_{\rm dmap}$    &                                               \multicolumn{3}{c|}{---}                                               &               4.75$\sigma$               &                  4.75$\sigma$                   &                  $4.75\sigma$                   \\ \hline
		\end{tabular}
	\end{center}
	\caption{Best-fit results of the MCMC analysis for $\Lambda$CDM involving the ACT data and pertinent likelihood combinations for reference.}
	\label{tab:bestfitLCDM}
\end{table}

\begin{table}
	\centering
	\renewcommand{\arraystretch}{1.45}
	\setlength\tabcolsep{0pt}
	\fontsize{8}{11}\selectfont
	\begin{tabular}{|>{\centering\arraybackslash}p{2.1cm}|>{\centering\arraybackslash}p{3.1cm}|>{\centering\arraybackslash}p{3.1cm}|>{\centering\arraybackslash}p{3.1cm}|}
		\hline
		\hspace*{.5cm}Name\hspace*{.5cm}                                & $\Lambda$CDM BICEP18   & NEDE: Base+BICEP18              & NEDE: Base + BICEP18 + S$H_0$ES \\ \hline\hline
		\multirow{2}{1.5cm}{\centering $\Omega_m$}                      & $0.311$                & $0.308$                         & $0.301$                         \\
		                                                                & $0.31072\pm 0.00546$   & $0.30818\pm 0.00587$            & $0.30006\pm 0.00518$            \\ \hline
		\multirow{2}{1.5cm}{\centering $\Omega_m h^2$}                  & $0.142$                & $0.147$                         & $0.153$                         \\
		                                                                & $0.142337\pm 0.000861$ & $0.14654^{+0.00196}_{-0.00435}$ & $0.15355\pm 0.00328$            \\ \hline
		\multirow{2}{1.5cm}{\centering $\Omega_\Lambda$ }               & $0.689$                & $0.692$                         & $0.699$                         \\
		                                                                & $0.68920\pm 0.00546$   & $0.69174\pm 0.00588$            & $0.69987\pm 0.00518$            \\ \hline
		\multirow{2}{1.5cm}{\centering $z_{\rm re}$}                    & $7.617$                & $7.958$                         & $7.670$                         \\
		                                                                & $7.904\pm 0.714$       & $7.906\pm 0.724$                & $8.079\pm 0.735$                \\ \hline
		\multirow{2}{1.5cm}{\centering $\sigma_8$}                      & $0.809$                & $0.825$                         & $0.836$                         \\
		                                                                & $0.81066\pm 0.00586$   & $0.82123^{+0.00853}_{-0.0121}$  & $0.84004\pm 0.00963$            \\ \hline
		\multirow{2}{1.5cm}{\centering $\sigma_8/h^{0.5}$ }             & $0.984$                & $0.993$                         & $0.990$                         \\
		                                                                & $0.98535\pm 0.00815$   & $0.98893\pm 0.00877$            & $0.99320\pm 0.00879$            \\ \hline
		\multirow{2}{1.5cm}{\centering $\sigma_8\Omega_{\rm m}^{0.5}$}  & $0.451$                & $0.458$                         & $0.459$                         \\
		                                                                & $0.45187\pm 0.00559$   & $0.45587\pm 0.00651$            & $0.46014\pm 0.00671$            \\ \hline
		\multirow{2}{1.5cm}{\centering $\sigma_8\Omega_{\rm m}^{0.25}$} & $0.604$                & $0.615$                         & $0.619$                         \\
		                                                                & $0.60523\pm 0.00553$   & $0.61185^{+0.00691}_{-0.00842}$ & $0.62171\pm 0.00770$            \\ \hline
		\multirow{2}{1.5cm}{\centering Age [Gyr]}                       & $13.789$               & $13.543$                        & $13.194$                        \\
		                                                                & $13.7864\pm 0.0197$    & $13.565^{+0.227}_{-0.0838}$     & $13.175\pm 0.138$               \\ \hline
		\multirow{2}{1.5cm}{\centering $r_s^{\rm d}$}                   & $147.277$              & $144.705$                       & $141.263$                       \\
		                                                                & $147.240\pm 0.228$     & $144.95^{+2.36}_{-0.891}$       & $141.10\pm 1.52$                \\ \hline\hline
		$\sum \chi^2$                                                   & 4342.37                        & 4339.68                        & 4341.49                         \\ \hline
	\end{tabular}

	\caption{Best-fit values, means and $1\sigma$ confidence intervals for some derived parameters from the MCMC samples involving BICEP18}
	\label{fig:meansDerivedBicep18}
\end{table}

\begin{figure}[htbp]
	\centering
	\includegraphics[clip, width=.74\textwidth]{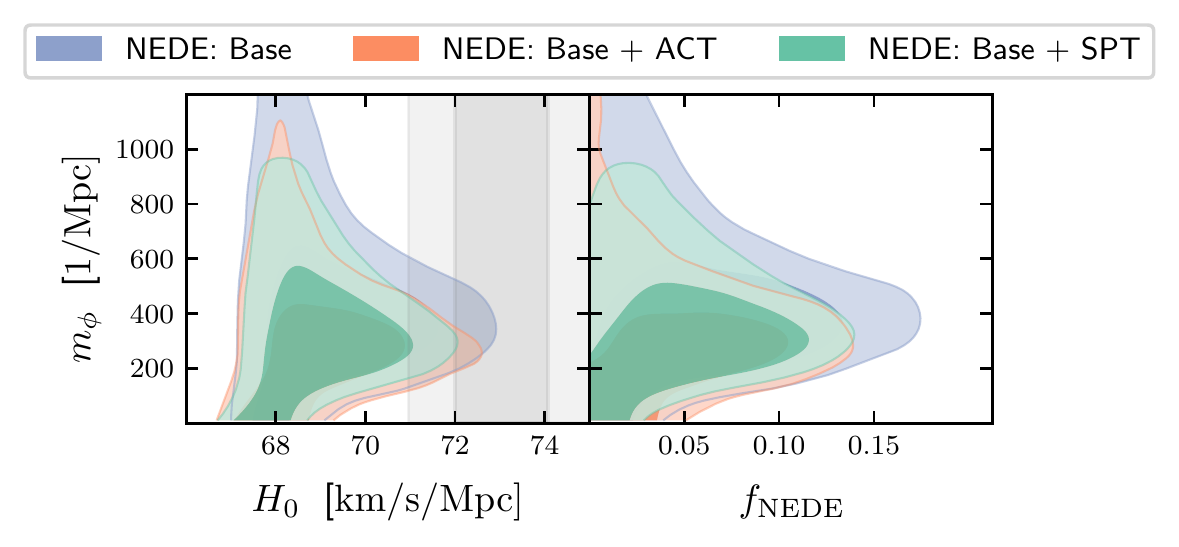}\\
	\includegraphics[clip, width=.99\textwidth]{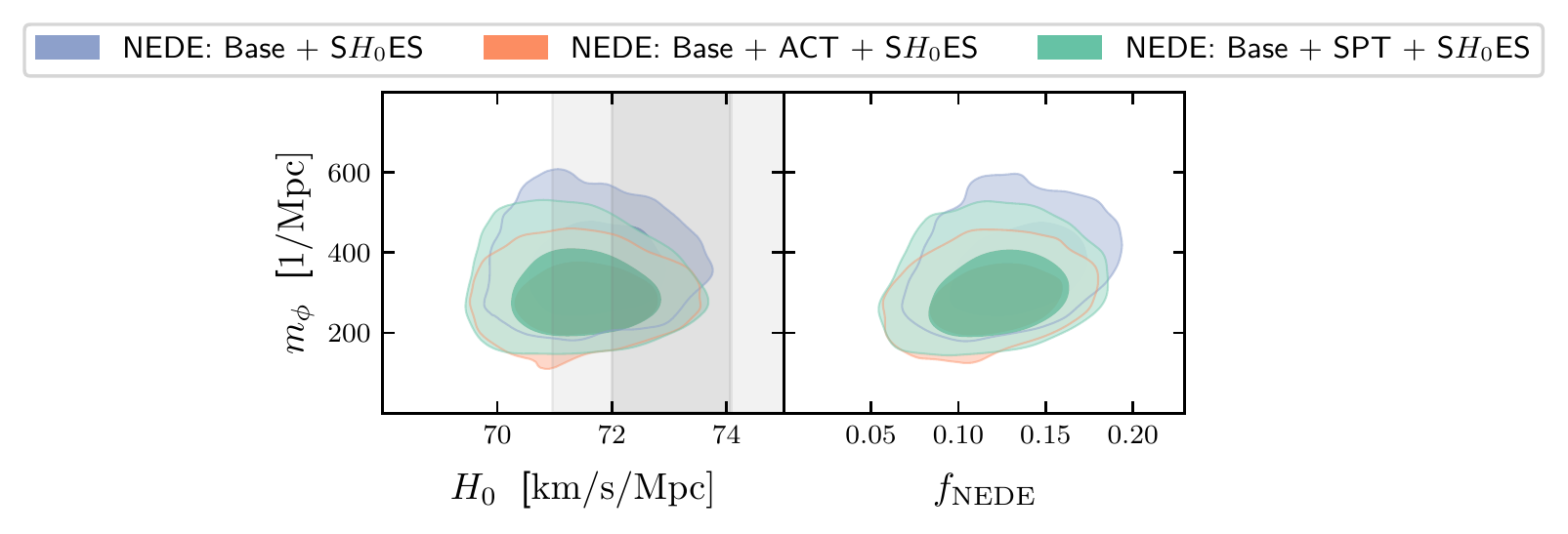}
	\caption{Mass posteriors from the different simulations of NEDE using the baseline datasets. Each plot displays the $68 \%$ and $95\%$ C.L. contours for the different dataset combinations. (Upper panels) MCMC runs without S$H_0$ES. (Lower panels) MCMC runs including the S$H_0$ES likelihood. }
	\label{fig:massPosteriors}
\end{figure}

\section[Individual chi2 for the different chains presented]{Individual $\chi^2$ for the different chains presented}

\begin{table}[htbp]
	\centering
	\renewcommand{\arraystretch}{1.5}
	\setlength\tabcolsep{5pt}
	\fontsize{8}{11}\selectfont
	\begin{tabular}{|l|c|}
		\hline
		\hspace*{0.5cm}Dataset\hspace*{4.3cm}   & \hspace*{1.2cm}$\chi^2$\hspace*{1.2cm} \\ \hline\hline
		Planck 2018\ lowl.TT                    &     $21.0697_{-0.5784}^{+0.5413} \quad (20.95)$      \\
		Planck 2018\ lowl.EE                    &     $397.0153_{-1.1716}^{+1.2501} \quad (395.92)$     \\
		Planck 2018\ lensing.clik               &     $10.1767_{-0.6600}^{+0.6169} \quad (10.08)$      \\
		Planck 2018\ highl\ plik.TTTEEE         &    $2361.91_{-6.3781}^{+6.3973} \quad (2341.16)$     \\
		BAO sdss\ dr7\ mgs                      &      $2.153_{-0.4515}^{+0.4509} \quad (2.155)$      \\
		BAO sixdf\ 2011\ bao                    &      $0.0366_{-0.0347}^{+0.0365} \quad (0.018)$      \\
		BAO sdss\ dr12\ consensus\ bao          &      $3.734_{-0.3147}^{+0.3269} \quad (3.426)$      \\
		SN Pantheon                             &    $1034.79_{-0.0516}^{+0.0528} \quad (1034.73))$     \\
		ACTPol\_lite DR4 for combining w Planck &     $242.20_{-4.0039}^{+4.0044} \quad (233.67)$     \\
		S$H_0$ES 2021                           &      $3.814_{-3.0405}^{+3.0903} \quad (2.188)$      \\ \hline
		\hfill$\Sigma$                          &                 $4076.9 \quad (4044.31)$				\\ \hline
	\end{tabular}
	\caption{Individual $\chi^2$ posteriors means and deviations with the best-fit in parenthesis, for the likelihoods used in the simulation of subsection~\ref{subsec:freeEOS} including ACT and sampling over $w\nede$.}
	\label{tab:chiSqFreeEOS}
\end{table}

\begin{table}[htbp]
	\begin{center}
		\renewcommand{\arraystretch}{1.12}
		\setlength\tabcolsep{0pt}
		\fontsize{7}{9}\selectfont

		\begin{tabular}{|l||c|c|c|c|c|c|}
			\hline
			\multirow{3}{1.5cm}{\hspace*{.2cm}Dataset} &                                                                                                          \multicolumn{6}{c|}{$\Lambda$CDM (Base = Planck+BAO+SN)}                                                                                                          \\
			\hhline{~|------|}                         & \multirow{2}{1.5cm}{\centering Base} & \multirow{2}{1.5cm}{\centering+ACT} & \multirow{2}{1.5cm}{\centering+SPT} & \multirow{2}{1.5cm}{\centering+S$H_0$ES21} & \multirow{2}{1.5cm}{\centering+ACT +S$H_0$ES21} & \multirow{2}{1.5cm}{\centering+SPT +S$H_0$ES21} \\
			                                           &                                      &                                     &                                     &                                            &                                                 &                                                  \\ \hline\hline
			\hspace*{.1cm}	Pl.18 lowl.TT               &                23.13                 &                22.63                &               22.836                &                   22.48                    &                      22.25                      &                      22.14                       \\
			\hspace*{.1cm}	Pl.18 lowl.EE               &                396.81                &               396.56                &               395.695               &                   397.69                   &                     397.21                      &                      397.48                      \\
			\hspace*{.1cm}	Pl.18 lensing.clik          &                 8.74                 &                8.73                 &                9.418                &                    9.18                    &                      8.62                       &                       8.95                       \\
			\hspace*{.1cm}	Pl.18 highl.TTTEEE$\;$      &               2337.97                &               2338.87               &              2340.448               &                  2342.27                   &                     2339.76                     &                     2342.11                      \\
			\hspace*{.1cm}	bao.sdss dr7 mgs            &                 1.21                 &                1.44                 &                1.340                &                    2.24                    &                      1.87                       &                       1.97                       \\
			\hspace*{.1cm}	bao.sixdf 2011 bao          &                 0.03                 &                0.01                 &                0.018                &                    0.03                    &                      0.00                       &                       0.01                       \\
			\hspace*{.1cm}	bao.sdss dr12 Cons.$\,$     &                 4.51                 &                3.89                 &                4.163                &                    3.46                    &                      3.38                       &                       3.39                       \\
			\hspace*{.1cm}	sn.pantheon                 &               1035.05                &               1034.92               &              1034.954               &                  1034.73                   &                     1034.77                     &                     1034.75                      \\
			\hspace*{.1cm}	ACTPol lite DR4             &                  --                  &               235.33                &                 --                  &                     --                     &                     235.79                      &                        --                        \\
			\hspace*{.1cm}   SPT3G Y1.TEEE             &                  --                  &                 --                  &              1117.716               &                     --                     &                       --                        &                     1118.68                      \\
			\hspace*{.1cm}	S$H_0$ES                    &                  --                  &                 --                  &                 --                  &                   17.98                    &                      21.28                      &                      20.24                       \\ \hline
			\hspace*{.1cm}	Total $\chi^2$              &               3807.45                &               4042.38               &              4926.589               &                  3830.06                   &                     4064.94                     &                     4949.71                      \\ \hline
			\hspace*{.1cm}	$Q_{\rm dmap}$              &                                             \multicolumn{3}{c|}{--}                                              &                4.75$\sigma$                &                  $4.75\sigma$                   &                   4.75$\sigma$                   \\ \hline
		\end{tabular}
	\end{center}
	\caption{$\chi^2$ values for the individual likelihoods used in the different MCMC analyses, together with the respective totals and $Q_{\rm dmap}$.}
	\label{tab:chiSqLCDM}
\end{table}

\begin{table}[htbp]
	\begin{center}
		\renewcommand{\arraystretch}{1.12}
		\setlength\tabcolsep{0pt}
		\fontsize{7}{9}\selectfont
		\begin{tabular}{|l||c|c|c|c|c|}
			\hline
			\multirow{3}{1.5cm}{\hspace*{.2cm}Dataset} &                                                                                                \multicolumn{5}{c|}{NEDE fixed EOS (Base = Planck+BAO+SN)}                                                                                                 \\
			\hhline{~|-----|}                          & \multirow{2}{1.5cm}{\centering Base} & \multirow{2}{1.5cm}{\centering+ACT} & \multirow{2}{1.5cm}{\centering+S$H_0$ES21} & \multirow{2}{1.5cm}{\centering+ACT +S$H_0$ES21} & \multirow{2}{1.5cm}{\centering+ACT $\;$fixed $m_\phi$} \\
			                                           &                                      &                                     &                                            &                                                 &                                                        \\ \hline\hline
			\hspace*{.1cm}	Pl.18 lowl.TT               &                21.69                 &                21.88                &                   20.73                    &                      20.51                      &                         21.03                          \\
			\hspace*{.1cm}	Pl.18 lowl.EE               &                396.09                &               395.71                &                   395.92                   &                     396.84                      &                         395.87                         \\
			\hspace*{.1cm}	Pl.18 lensing.clik          &                 9.55                 &                9.30                 &                    9.83                    &                      10.39                      &                          9.94                          \\
			\hspace*{.1cm}	Pl.18 highl.TTTEEE$\;$      &               2336.68                &               2338.11               &                  2338.51                   &                     2340.95                     &                        2337.23                         \\
			\hspace*{.1cm}	bao.sdss dr7 mgs            &                 1.47                 &                1.41                 &                    2.05                    &                      2.28                       &                          1.91                          \\
			\hspace*{.1cm}	bao.sixdf 2011 bao          &                 0.01                 &                0.01                 &                    0.01                    &                      0.03                       &                          0.00                          \\
			\hspace*{.1cm}	bao.sdss dr12 Cons.$\,$     &                 3.92                 &                3.99                 &                    3.44                    &                      3.51                       &                          3.41                          \\
			\hspace*{.1cm}	sn.pantheon                 &               1034.88                &               1034.92               &                  1034.73                   &                     1034.74                     &                        1034.75                         \\
			\hspace*{.1cm}	ACTPol lite DR4             &                  --                  &               235.21                &                     --                     &                     238.97                      &                         235.11                         \\
			\hspace*{.1cm}	S$H_0$ES                    &                  --                  &                 --                  &                    1.52                    &                      0.84                       &                           --                           \\ \hline
			\hspace*{.1cm}	Total $\chi^2$              &               3804.26                &               4040.56               &                  3806.74                   &                     4049.05                     &                        4039.26                         \\ \hline
			\hspace*{.1cm}	$\Delta\chi^2$              &               $-3.19$                &    \cellcolor{violet!40}$-1.82$     &        \cellcolor{teal!40} $-23.32$        &           \cellcolor{red!40} $-15.89$           &              \cellcolor{green!40} $-3.13$              \\ \hline
			\hspace*{.1cm}	$Q_{\rm dmap}$              &                          \multicolumn{2}{c|}{--}                           &                1.57$\sigma$                &                   2.9$\sigma$                   &                           --                           \\ \hline
		\end{tabular}
	\end{center}
	\caption{$\chi^2$ values for the individual likelihoods used in the different MCMC analysis involving ACT and corresponding reference runs, together with the respective totals and $Q_{\rm dmap}$.}
	\label{tab:chiSqACT}
\end{table}

\begin{table}
	\begin{center}
		\renewcommand{\arraystretch}{1.12}
		\setlength\tabcolsep{0pt}
		\fontsize{7}{9}\selectfont
		\begin{tabular}{|l||c|c|c|c|c|}
			\hline
			\multirow{3}{1.5cm}{\hspace*{.2cm}Dataset} &                                                                                     \multicolumn{5}{c|}{NEDE fixed EOS (Base = Planck+BAO+SN)}                                                                                     \\
			\hhline{~|-----|}                          & \multirow{2}{1.5cm}{\centering Base} & \multirow{2}{1.5cm}{\centering+SPT} & \multirow{2}{1.5cm}{\centering+S$H_0$ES21} & \multirow{2}{1.5cm}{\centering+SPT +S$H_0$ES21} & \multirow{2}{1.5cm}{\centering+SPT $\;$fixed $m_\phi$} \\
			                                           &                                      &                                     &                                            &                                                 &                                                        \\ \hline\hline
			\hspace*{.1cm}   Pl.18 lowl.TT             &                21.686                &               21.664                &                   20.727                   &                     20.749                      &                         21.725                         \\
			\hspace*{.1cm}   Pl.18 lowl.EE             &               396.087                &               396.166               &                  395.918                   &                     397.283                     &                        396.445                         \\
			\hspace*{.1cm}	 Pl.18 lensing.clik         &                9.545                 &                9.314                &                   9.834                    &                      9.851                      &                         9.234                          \\
			\hspace*{.1cm}	 Pl.18 highl.TTTEEE$\;$     &               2336.679               &              2337.241               &                  2338.514                  &                    2337.810                     &                        2337.021                        \\
			\hspace*{.1cm}   bao.sdss dr7 mgs          &                1.465                 &                1.409                &                   2.045                    &                      2.331                      &                         1.526                          \\
			\hspace*{.1cm}	 bao.sixdf 2011 bao         &                0.008                 &                0.012                &                   0.010                    &                      0.036                      &                         0.005                          \\
			\hspace*{.1cm}	 bao.sdss dr12 Cons.        &                3.918                 &                4.045                &                   3.441                    &                      3.564                      &                         3.814                          \\
			\hspace*{.1cm}	 sn.pantheon                &               1034.876               &              1034.901               &                  1034.735                  &                    1034.745                     &                        1034.848                        \\
			\hspace*{.1cm}   SPT3G Y1.TEEE             &                  --                  &              1118.515               &                     --                     &                    1118.718                     &                        1118.607                        \\
			\hspace*{.1cm}	 S$H_0$ES                   &                  --                  &                 --                  &                   1.517                    &                      1.494                      &                           --                           \\ \hline
			\hspace*{.1cm}   Total chi2                &               3804.265               &              4923.266               &                  3806.741                  &                    4926.580                     &                        4923.224                        \\ \hline
			\hspace*{.1cm}	$\Delta\chi^2$              &                -3.19                 &    \cellcolor{violet!40}  -3.32     &         \cellcolor{teal!40} -23.32         &            \cellcolor{red!40} -23.13            &             \cellcolor{green!40}    -3.37              \\ \hline
			$Q_{\rm dmap}$                             &                                      &                                     &                $1.57\sigma$                &                  $1.82\sigma$                   &                                                        \\ \hline
		\end{tabular}
	\end{center}
	\caption{$\chi^2$ values for the individual likelihoods used in the different MCMC analysis involving SPT and corresponding reference runs, together with the respective totals and $Q_{\rm dmap}$.}
	\label{tab:chiSqSPT}
\end{table}

\begin{table}[htbp]
	\begin{center}
		\renewcommand{\arraystretch}{1.45}
		\setlength\tabcolsep{0pt}
		\fontsize{8}{11}\selectfont
		\begin{tabular}{|>{\centering\arraybackslash}p{3.1cm}||>{\centering\arraybackslash}p{2.1cm}|>{\centering\arraybackslash}p{2.1cm}|>{\centering\arraybackslash}p{2.1cm}|}
			\hline
			\multirow{2}{3cm}{\hspace*{.2cm}{\bf Dataset }} & \multirow{2}{1.7cm}{LCDM: Base + BICEP18} & \multirow{2}{1.7cm}{NEDE: Base + BICEP18} & \multirow{2}{1.7cm}{NEDE: Base + BICEP18 + S$H_0$ES} \\
			&                                           &                                           &                                                      \\ \hline\hline
			Planck18 low $\ell$ TT                  & 23.25                                     & 22.16                                     & 20.91                                                \\
			Planck18 low $\ell$ EE                  & 395.99                                    & 396.51                                    & 395.87                                            \\
			Planck18 high $\ell$ plik TTTEEE        & 2338.74                                   & 2336.64                                   & 2337.51                                            \\
			Planck18 lensing clik                   & 8.84                                      & 9.09                                      & 9.71                                            \\
			BAO sdss dr7 mgs                        & 1.26                                      & 1.41                                      & 1.93                                            \\
			BAO sixdf 2011 bao                      & 0.028                                     & 0.012                                     & 0.004                                          \\
			BAO sdss dr12 consensus                 & 4.37                                      & 4.03                                      & 3.45                                            \\
			SN Pantheon                             & 1035.01                                   & 1034.90                                   & 1034.74                                            \\
			BICEP/Keck 2018                         & 534.88                                    & 534.93                                    & 534.80                                            \\
			S$H_0$ES21                              & --                                        & --                                        & 2.57                                            \\ \hline
			$\sum \chi^2$                           & 4342.37                                   & 4339.68                                   & 4341.49                                            \\ \hline
		\end{tabular}
	\end{center}
	\caption{Individual and total $\chi^2$ values for MCMC runs involving the BICEP18 dataset described in subsection \ref{subsec:bicep}.}
	\label{tab:chiSqBicep}
\end{table}

\bibliography{references}

\providecommand{\href}[2]{#2}\begingroup\raggedright\begin{thebibliography}{10}

\bibitem{Nadathur:2020kvq}
S.~Nadathur, W.J.~Percival, F.~Beutler and H.A.~Winther, \emph{Testing
  {{Low-Redshift Cosmic Acceleration}} with {{Large-Scale Structure}}},
  \href{https://doi.org/10.1103/PhysRevLett.124.221301}{\emph{Phys. Rev. Lett.}
  {\bfseries 124} (2020) 221301}.

\bibitem{Abdalla:2022yfr}
E.~Abdalla et~al., \emph{{Cosmology intertwined: A review of the particle
  physics, astrophysics, and cosmology associated with the cosmological
  tensions and anomalies}},
  \href{https://doi.org/10.1016/j.jheap.2022.04.002}{\emph{JHEAp} {\bfseries
  34} (2022) 49} [\href{https://arxiv.org/abs/2203.06142}{{\ttfamily
  2203.06142}}].

\bibitem{Knox:2019rjx}
L.~Knox and M.~Millea, \emph{{Hubble constant hunter's guide}},
  \href{https://doi.org/10.1103/PhysRevD.101.043533}{\emph{Phys. Rev. D}
  {\bfseries 101} (2020) 043533}
  [\href{https://arxiv.org/abs/1908.03663}{{\ttfamily 1908.03663}}].

\bibitem{DiValentino:2020zio}
E.~Di~Valentino et~al., \emph{{Snowmass2021 - Letter of interest cosmology
  intertwined II: The hubble constant tension}},
  \href{https://doi.org/10.1016/j.astropartphys.2021.102605}{\emph{Astropart.
  Phys.} {\bfseries 131} (2021) 102605}
  [\href{https://arxiv.org/abs/2008.11284}{{\ttfamily 2008.11284}}].

\bibitem{Planck:2018vyg}
{\scshape Planck} collaboration, \emph{{Planck 2018 results. VI. Cosmological
  parameters}},
  \href{https://doi.org/10.1051/0004-6361/201833910}{\emph{Astron. Astrophys.}
  {\bfseries 641} (2020) A6}
  [\href{https://arxiv.org/abs/1807.06209}{{\ttfamily 1807.06209}}].

\bibitem{Riess:2021jrx}
A.G.~Riess et~al., \emph{{A Comprehensive Measurement of the Local Value of the
  Hubble Constant with 1 km s{$^{-1}$} Mpc{$^{-1}$} Uncertainty from the Hubble
  Space Telescope and the SH0ES Team}},
  \href{https://doi.org/10.3847/2041-8213/ac5c5b}{\emph{Astrophys. J. Lett.}
  {\bfseries 934} (2022) L7}
  [\href{https://arxiv.org/abs/2112.04510}{{\ttfamily 2112.04510}}].

\bibitem{Wojtak:2013gda}
R.~Wojtak, A.~Knebe, W.A.~Watson, I.T.~Iliev, S.~He{\ss}, D.~Rapetti et~al.,
  \emph{Cosmic variance of the local {{Hubble}} flow in large-scale
  cosmological simulations},
  \href{https://doi.org/10.1093/mnras/stt2321}{\emph{Mon. Not. R. Astron Soc.}
  {\bfseries 438} (2014) 1805}.

\bibitem{Odderskov:2017ivg}
I.~Odderskov, S.~Hannestad and J.~Brandbyge, \emph{The variance of the locally
  measured {{Hubble}} parameter explained with different estimators},
  \href{https://doi.org/10.1088/1475-7516/2017/03/022}{\emph{J. Cosmol.
  Astropart. Phys.} {\bfseries 2017} (2017) 022}.

\bibitem{Wu:2017fpr}
H.-Y.~Wu and D.~Huterer, \emph{Sample variance in the local measurements of the
  {{Hubble}} constant}, \href{https://doi.org/10.1093/mnras/stx1967}{\emph{Mon.
  Not. R. Astron Soc.} {\bfseries 471} (2017) 4946}.

\bibitem{Davis:2019wet}
T.M.~Davis, S.R.~Hinton, C.~Howlett and J.~Calcino, \emph{Can redshift errors
  bias measurements of the {{Hubble Constant}}?},
  \href{https://doi.org/10.1093/mnras/stz2652}{\emph{Mon. Not. R. Astron Soc.}
  {\bfseries 490} (2019) 2948}.

\bibitem{DiValentino:2021izs}
E.~Di~Valentino, O.~Mena, S.~Pan, L.~Visinelli, W.~Yang, A.~Melchiorri et~al.,
  \emph{{In the realm of the Hubble tension\textemdash{}a review of
  solutions}}, \href{https://doi.org/10.1088/1361-6382/ac086d}{\emph{Class.
  Quant. Grav.} {\bfseries 38} (2021) 153001}
  [\href{https://arxiv.org/abs/2103.01183}{{\ttfamily 2103.01183}}].

\bibitem{Schoneberg:2021qvd}
N.~Sch\"oneberg, G.~Franco~Abell\'an, A.~P\'erez~S\'anchez, S.J.~Witte,
  V.~Poulin and J.~Lesgourgues, \emph{{The $H_0$ Olympics: A fair ranking of
  proposed models}},
  \href{https://doi.org/10.1016/j.physrep.2022.07.001}{\emph{Phys. Rept.}
  {\bfseries 984} (2022) 1} [\href{https://arxiv.org/abs/2107.10291}{{\ttfamily
  2107.10291}}].

\bibitem{Bernal:2016gxb}
J.L.~Bernal, L.~Verde and A.G.~Riess, \emph{{The trouble with $H_0$}},
  \href{https://doi.org/10.1088/1475-7516/2016/10/019}{\emph{J. Cosmol.
  Astropart. Phys.} {\bfseries 10} (2016) 019}
  [\href{https://arxiv.org/abs/1607.05617}{{\ttfamily 1607.05617}}].

\bibitem{Aylor:2018drw}
K.~Aylor, M.~Joy, L.~Knox, M.~Millea, S.~Raghunathan and W.L.K.~Wu,
  \emph{{Sounds Discordant: Classical Distance Ladder \& $\Lambda$CDM -based
  Determinations of the Cosmological Sound Horizon}},
  \href{https://doi.org/10.3847/1538-4357/ab0898}{\emph{Astrophys. J.}
  {\bfseries 874} (2019) 4} [\href{https://arxiv.org/abs/1811.00537}{{\ttfamily
  1811.00537}}].

\bibitem{Arendse:2019hev}
N.~Arendse et~al., \emph{{Cosmic dissonance: are new physics or systematics
  behind a short sound horizon?}},
  \href{https://doi.org/10.1051/0004-6361/201936720}{\emph{Astron. Astrophys.}
  {\bfseries 639} (2020) A57}
  [\href{https://arxiv.org/abs/1909.07986}{{\ttfamily 1909.07986}}].

\bibitem{DiValentino:2020vvd}
E.~Di~Valentino et~al., \emph{{Cosmology intertwined III: $f\sigma_8$ and
  $S_8$}},
  \href{https://doi.org/10.1016/j.astropartphys.2021.102604}{\emph{Astropart.
  Phys.} {\bfseries 131} (2021) 102604}
  [\href{https://arxiv.org/abs/2008.11285}{{\ttfamily 2008.11285}}].

\bibitem{Nunes:2021ipq}
R.C.~Nunes and S.~Vagnozzi, \emph{{Arbitrating the S8 discrepancy with growth
  rate measurements from redshift-space distortions}},
  \href{https://doi.org/10.1093/mnras/stab1613}{\emph{Mon. Not. Roy. Astron.
  Soc.} {\bfseries 505} (2021) 5427}
  [\href{https://arxiv.org/abs/2106.01208}{{\ttfamily 2106.01208}}].

\bibitem{Asadi:2022njl}
P.~Asadi et~al., \emph{{Early-Universe Model Building}},
  \href{https://arxiv.org/abs/2203.06680}{{\ttfamily 2203.06680}}.

\bibitem{Joudaki:2016kym}
S.~Joudaki et~al., \emph{{KiDS-450: Testing extensions to the standard
  cosmological model}}, \href{https://doi.org/10.1093/mnras/stx998}{\emph{Mon.
  Not. Roy. Astron. Soc.} {\bfseries 471} (2017) 1259}
  [\href{https://arxiv.org/abs/1610.04606}{{\ttfamily 1610.04606}}].

\bibitem{Hildebrandt:2016iqg}
H.~Hildebrandt et~al., \emph{{KiDS-450: Cosmological parameter constraints from
  tomographic weak gravitational lensing}},
  \href{https://doi.org/10.1093/mnras/stw2805}{\emph{Mon. Not. Roy. Astron.
  Soc.} {\bfseries 465} (2017) 1454}
  [\href{https://arxiv.org/abs/1606.05338}{{\ttfamily 1606.05338}}].

\bibitem{Hildebrandt:2018yau}
H.~Hildebrandt et~al., \emph{{KiDS+VIKING-450: Cosmic shear tomography with
  optical and infrared data}},
  \href{https://doi.org/10.1051/0004-6361/201834878}{\emph{Astron. Astrophys.}
  {\bfseries 633} (2020) A69}
  [\href{https://arxiv.org/abs/1812.06076}{{\ttfamily 1812.06076}}].

\bibitem{KiDS:2020suj}
{\scshape KiDS} collaboration, \emph{{KiDS-1000 Cosmology: Cosmic shear
  constraints and comparison between two point statistics}},
  \href{https://doi.org/10.1051/0004-6361/202039070}{\emph{Astron. Astrophys.}
  {\bfseries 645} (2021) A104}
  [\href{https://arxiv.org/abs/2007.15633}{{\ttfamily 2007.15633}}].

\bibitem{Poulin:2018dzj}
V.~Poulin, T.L.~Smith, D.~Grin, T.~Karwal and M.~Kamionkowski,
  \emph{Cosmological implications of ultralight axionlike fields},
  \href{https://doi.org/10.1103/PhysRevD.98.083525}{\emph{Phys. Rev. D}
  {\bfseries 98} (2018) 083525}.

\bibitem{DAmico:2020ods}
G.~D'Amico, L.~Senatore, P.~Zhang and H.~Zheng, \emph{{The Hubble Tension in
  Light of the Full-Shape Analysis of Large-Scale Structure Data}},
  \href{https://doi.org/10.1088/1475-7516/2021/05/072}{\emph{J. Cosmol.
  Astropart. Phys.} {\bfseries 05} (2021) 072}
  [\href{https://arxiv.org/abs/2006.12420}{{\ttfamily 2006.12420}}].

\bibitem{Niedermann:2020dwg}
F.~Niedermann and M.S.~Sloth, \emph{Resolving the {{Hubble}} tension with new
  early dark energy},
  \href{https://doi.org/10.1103/PhysRevD.102.063527}{\emph{Phys. Rev. D}
  {\bfseries 102} (2020) 063527}.

\bibitem{Niedermann:2020qbw}
F.~Niedermann and M.S.~Sloth, \emph{{New Early Dark Energy is compatible with
  current LSS data}},
  \href{https://doi.org/10.1103/PhysRevD.103.103537}{\emph{Phys. Rev. D}
  {\bfseries 103} (2021) 103537}
  [\href{https://arxiv.org/abs/2009.00006}{{\ttfamily 2009.00006}}].

\bibitem{Murgia:2020ryi}
R.~Murgia, G.F.~Abell\'an and V.~Poulin, \emph{{Early dark energy resolution to
  the Hubble tension in light of weak lensing surveys and lensing anomalies}},
  \href{https://doi.org/10.1103/PhysRevD.103.063502}{\emph{Phys. Rev. D}
  {\bfseries 103} (2021) 063502}
  [\href{https://arxiv.org/abs/2009.10733}{{\ttfamily 2009.10733}}].

\bibitem{Smith:2020rxx}
T.L.~Smith, V.~Poulin, J.L.~Bernal, K.K.~Boddy, M.~Kamionkowski and R.~Murgia,
  \emph{{Early dark energy is not excluded by current large-scale structure
  data}}, \href{https://doi.org/10.1103/PhysRevD.103.123542}{\emph{Phys. Rev.
  D} {\bfseries 103} (2021) 123542}
  [\href{https://arxiv.org/abs/2009.10740}{{\ttfamily 2009.10740}}].

\bibitem{Simon:2022adh}
T.~Simon, P.~Zhang, V.~Poulin and T.L.~Smith, \emph{{Updated constraints from
  the effective field theory analysis of BOSS power spectrum on Early Dark
  Energy}},  \href{https://arxiv.org/abs/2208.05930}{{\ttfamily 2208.05930}}.

\bibitem{Niedermann:2019olb}
F.~Niedermann and M.S.~Sloth, \emph{New early dark energy},
  \href{https://doi.org/10.1103/PhysRevD.103.L041303}{\emph{Phys. Rev. D}
  {\bfseries 103} (2021) L041303}.

\bibitem{Allali:2021azp}
I.J.~Allali, M.P.~Hertzberg and F.~Rompineve, \emph{{Dark sector to restore
  cosmological concordance}},
  \href{https://doi.org/10.1103/PhysRevD.104.L081303}{\emph{Phys. Rev. D}
  {\bfseries 104} (2021) L081303}
  [\href{https://arxiv.org/abs/2104.12798}{{\ttfamily 2104.12798}}].

\bibitem{Freese:2021rjq}
K.~Freese and M.W.~Winkler, \emph{Chain {{Early Dark Energy}}: {{Solving}} the
  {{Hubble Tension}} and {{Explaining Today}}'s {{Dark Energy}}},
  \href{https://doi.org/10.1103/PhysRevD.104.083533}{\emph{Phys. Rev. D}
  {\bfseries 104} (2021) 083533}
  [\href{https://arxiv.org/abs/2102.13655}{{\ttfamily 2102.13655}}].

\bibitem{Niedermann:2021vgd}
F.~Niedermann and M.S.~Sloth, \emph{{Hot new early dark energy}},
  \href{https://doi.org/10.1103/PhysRevD.105.063509}{\emph{Phys. Rev. D}
  {\bfseries 105} (2022) 063509}
  [\href{https://arxiv.org/abs/2112.00770}{{\ttfamily 2112.00770}}].

\bibitem{Niedermann:2021ijp}
F.~Niedermann and M.S.~Sloth, \emph{Hot {{New Early Dark Energy}}: {{Towards}}
  a {{Unified Dark Sector}} of {{Neutrinos}}, {{Dark Energy}} and {{Dark
  Matter}}},  \href{https://arxiv.org/abs/2112.00759}{{\ttfamily 2112.00759}}.

\bibitem{Karwal:2016vyq}
T.~Karwal and M.~Kamionkowski, \emph{{Dark energy at early times, the Hubble
  parameter, and the string axiverse}},
  \href{https://doi.org/10.1103/PhysRevD.94.103523}{\emph{Phys. Rev. D}
  {\bfseries 94} (2016) 103523}
  [\href{https://arxiv.org/abs/1608.01309}{{\ttfamily 1608.01309}}].

\bibitem{Poulin:2018cxd}
V.~Poulin, T.L.~Smith, T.~Karwal and M.~Kamionkowski, \emph{Early {{Dark
  Energy}} can {{Resolve}} the {{Hubble Tension}}},
  \href{https://doi.org/10.1103/PhysRevLett.122.221301}{\emph{Phys. Rev. Lett.}
  {\bfseries 122} (2019) 221301}.

\bibitem{Smith:2019ihp}
T.L.~Smith, V.~Poulin and M.A.~Amin, \emph{{Oscillating scalar fields and the
  Hubble tension: a resolution with novel signatures}},
  \href{https://doi.org/10.1103/PhysRevD.101.063523}{\emph{Phys. Rev. D}
  {\bfseries 101} (2020) 063523}
  [\href{https://arxiv.org/abs/1908.06995}{{\ttfamily 1908.06995}}].

\bibitem{Poulin:2021bjr}
V.~Poulin, T.L.~Smith and A.~Bartlett, \emph{{Dark energy at early times and
  ACT data: A larger Hubble constant without late-time priors}},
  \href{https://doi.org/10.1103/PhysRevD.104.123550}{\emph{Phys. Rev. D}
  {\bfseries 104} (2021) 123550}
  [\href{https://arxiv.org/abs/2109.06229}{{\ttfamily 2109.06229}}].

\bibitem{LaPosta:2021pgm}
A.~La~Posta, T.~Louis, X.~Garrido and J.C.~Hill, \emph{{Constraints on
  prerecombination early dark energy from SPT-3G public data}},
  \href{https://doi.org/10.1103/PhysRevD.105.083519}{\emph{Phys. Rev. D}
  {\bfseries 105} (2022) 083519}
  [\href{https://arxiv.org/abs/2112.10754}{{\ttfamily 2112.10754}}].

\bibitem{Hill:2021yec}
J.C.~Hill et~al., \emph{The {{Atacama Cosmology Telescope}}: {{Constraints}} on
  {{Pre-Recombination Early Dark Energy}}},
  \href{https://doi.org/10.1103/PhysRevD.105.123536}{\emph{Phys. Rev. D}
  {\bfseries 105} (2022) 123536}
  [\href{https://arxiv.org/abs/2109.04451}{{\ttfamily 2109.04451}}].

\bibitem{Smith:2022hwi}
T.L.~Smith, M.~Lucca, V.~Poulin, G.F.~Abellan, L.~Balkenhol, K.~Benabed et~al.,
  \emph{{Hints of early dark energy in Planck, SPT, and ACT data: New physics
  or systematics?}},
  \href{https://doi.org/10.1103/PhysRevD.106.043526}{\emph{Phys. Rev. D}
  {\bfseries 106} (2022) 043526}
  [\href{https://arxiv.org/abs/2202.09379}{{\ttfamily 2202.09379}}].

\bibitem{Jiang:2022uyg}
J.-Q.~Jiang and Y.-S.~Piao, \emph{Toward early dark energy and ns=1 with
  {{Planck}}, {{ACT}}, and {{SPT}} observations},
  \href{https://doi.org/10.1103/PhysRevD.105.103514}{\emph{Phys. Rev. D}
  {\bfseries 105} (2022) 103514}.

\bibitem{DAmico:2021fhz}
G.~D'Amico, N.~Kaloper and A.~Westphal, \emph{{General double monodromy
  inflation}}, \href{https://doi.org/10.1103/PhysRevD.105.103527}{\emph{Phys.
  Rev. D} {\bfseries 105} (2022) 103527}
  [\href{https://arxiv.org/abs/2112.13861}{{\ttfamily 2112.13861}}].

\bibitem{Enqvist:2001zp}
K.~Enqvist and M.S.~Sloth, \emph{{Adiabatic CMB perturbations in pre - big bang
  string cosmology}},
  \href{https://doi.org/10.1016/S0550-3213(02)00043-3}{\emph{Nucl. Phys. B}
  {\bfseries 626} (2002) 395}
  [\href{https://arxiv.org/abs/hep-ph/0109214}{{\ttfamily hep-ph/0109214}}].

\bibitem{Lyth:2001nq}
D.H.~Lyth and D.~Wands, \emph{{Generating the curvature perturbation without an
  inflaton}}, \href{https://doi.org/10.1016/S0370-2693(01)01366-1}{\emph{Phys.
  Lett. B} {\bfseries 524} (2002) 5}
  [\href{https://arxiv.org/abs/hep-ph/0110002}{{\ttfamily hep-ph/0110002}}].

\bibitem{Moroi:2001ct}
T.~Moroi and T.~Takahashi, \emph{{Effects of cosmological moduli fields on
  cosmic microwave background}},
  \href{https://doi.org/10.1016/S0370-2693(01)01295-3}{\emph{Phys. Lett. B}
  {\bfseries 522} (2001) 215}
  [\href{https://arxiv.org/abs/hep-ph/0110096}{{\ttfamily hep-ph/0110096}}].

\bibitem{Starobinsky:1980te}
A.A.~Starobinsky, \emph{{A New Type of Isotropic Cosmological Models Without
  Singularity}},
  \href{https://doi.org/10.1016/0370-2693(80)90670-X}{\emph{Phys. Lett. B}
  {\bfseries 91} (1980) 99}.

\bibitem{Gouttenoire:2021jhk}
Y.~Gouttenoire, G.~Servant and P.~Simakachorn, \emph{{Kination cosmology from
  scalar fields and gravitational-wave signatures}},
  \href{https://arxiv.org/abs/2111.01150}{{\ttfamily 2111.01150}}.

\bibitem{Lin:2019qug}
M.-X.~Lin, G.~Benevento, W.~Hu and M.~Raveri, \emph{{Acoustic Dark Energy:
  Potential Conversion of the Hubble Tension}},
  \href{https://doi.org/10.1103/PhysRevD.100.063542}{\emph{Phys. Rev. D}
  {\bfseries 100} (2019) 063542}
  [\href{https://arxiv.org/abs/1905.12618}{{\ttfamily 1905.12618}}].

\bibitem{Vagnozzi:2021gjh}
S.~Vagnozzi, \emph{{Consistency tests of \ensuremath{\Lambda}CDM from the early
  integrated Sachs-Wolfe effect: Implications for early-time new physics and
  the Hubble tension}},
  \href{https://doi.org/10.1103/PhysRevD.104.063524}{\emph{Phys. Rev. D}
  {\bfseries 104} (2021) 063524}
  [\href{https://arxiv.org/abs/2105.10425}{{\ttfamily 2105.10425}}].

\bibitem{Ma:1995ey}
C.-P.~Ma and E.~Bertschinger, \emph{{Cosmological perturbation theory in the
  synchronous and conformal Newtonian gauges}},
  \href{https://doi.org/10.1086/176550}{\emph{Astrophys. J.} {\bfseries 455}
  (1995) 7} [\href{https://arxiv.org/abs/astro-ph/9506072}{{\ttfamily
  astro-ph/9506072}}].

\bibitem{Linde:1990gz}
A.D.~Linde, \emph{{Eternal extended inflation and graceful exit from old
  inflation without Jordan-Brans-Dicke}},
  \href{https://doi.org/10.1016/0370-2693(90)90521-7}{\emph{Phys. Lett. B}
  {\bfseries 249} (1990) 18}.

\bibitem{Adams:1990ds}
F.C.~Adams and K.~Freese, \emph{{Double field inflation}},
  \href{https://doi.org/10.1103/PhysRevD.43.353}{\emph{Phys. Rev. D} {\bfseries
  43} (1991) 353} [\href{https://arxiv.org/abs/hep-ph/0504135}{{\ttfamily
  hep-ph/0504135}}].

\bibitem{Copeland:1994vg}
E.J.~Copeland, A.R.~Liddle, D.H.~Lyth, E.D.~Stewart and D.~Wands, \emph{False
  vacuum inflation with {{Einstein}} gravity},
  \href{https://doi.org/10.1103/PhysRevD.49.6410}{\emph{Phys. Rev. D}
  {\bfseries 49} (1994) 6410}
  [\href{https://arxiv.org/abs/astro-ph/9401011}{{\ttfamily
  astro-ph/9401011}}].

\bibitem{Blas:2011rf}
D.~Blas, J.~Lesgourgues and T.~Tram, \emph{{The Cosmic Linear Anisotropy
  Solving System (CLASS) II: Approximation schemes}},
  \href{https://doi.org/10.1088/1475-7516/2011/07/034}{\emph{J. Cosmol.
  Astropart. Phys.} {\bfseries 07} (2011) 034}
  [\href{https://arxiv.org/abs/1104.2933}{{\ttfamily 1104.2933}}].

\bibitem{Torrado:2020dgo}
J.~Torrado and A.~Lewis, \emph{Cobaya: Code for {{Bayesian}} analysis of
  hierarchical physical models},
  \href{https://doi.org/10.1088/1475-7516/2021/05/057}{\emph{J. Cosmol.
  Astropart. Phys.} {\bfseries 05} (2021) 057}
  [\href{https://arxiv.org/abs/2005.05290}{{\ttfamily 2005.05290}}].

\bibitem{Gelman:1992zz}
A.~Gelman and D.B.~Rubin, \emph{{Inference from Iterative Simulation Using
  Multiple Sequences}},
  \href{https://doi.org/10.1214/ss/1177011136}{\emph{Statist. Sci.} {\bfseries
  7} (1992) 457}.

\bibitem{Planck:2019nip}
{\scshape Planck} collaboration, \emph{{Planck 2018 results. V. CMB power
  spectra and likelihoods}},
  \href{https://doi.org/10.1051/0004-6361/201936386}{\emph{Astron. Astrophys.}
  {\bfseries 641} (2020) A5}
  [\href{https://arxiv.org/abs/1907.12875}{{\ttfamily 1907.12875}}].

\bibitem{Beutler:2011hx}
F.~Beutler, C.~Blake, M.~Colless, D.H.~Jones, L.~Staveley-Smith, L.~Campbell
  et~al., \emph{The {{6dF Galaxy Survey}}: Baryon acoustic oscillations and the
  local {{Hubble}} constant},
  \href{https://doi.org/10.1111/j.1365-2966.2011.19250.x}{\emph{Mon. Not. R.
  Astron Soc.} {\bfseries 416} (2011) 3017}.

\bibitem{Ross:2014qpa}
A.J.~Ross, L.~Samushia, C.~Howlett, W.J.~Percival, A.~Burden and M.~Manera,
  \emph{The clustering of the {SDSS DR7} main galaxy sample \textendash{} {I.
  A} 4 per cent distance measure at $z = 0.15$},
  \href{https://doi.org/10.1093/mnras/stv154}{\emph{Mon. Not. Roy. Astron.
  Soc.} {\bfseries 449} (2015) 835}
  [\href{https://arxiv.org/abs/1409.3242}{{\ttfamily 1409.3242}}].

\bibitem{BOSS:2016wmc}
{\scshape BOSS} collaboration, \emph{{The clustering of galaxies in the
  completed SDSS-III Baryon Oscillation Spectroscopic Survey: cosmological
  analysis of the DR12 galaxy sample}},
  \href{https://doi.org/10.1093/mnras/stx721}{\emph{Mon. Not. Roy. Astron.
  Soc.} {\bfseries 470} (2017) 2617}
  [\href{https://arxiv.org/abs/1607.03155}{{\ttfamily 1607.03155}}].

\bibitem{Pan-STARRS1:2017jku}
{\scshape Pan-STARRS1} collaboration, \emph{{The Complete Light-curve Sample of
  Spectroscopically Confirmed SNe Ia from Pan-STARRS1 and Cosmological
  Constraints from the Combined Pantheon Sample}},
  \href{https://doi.org/10.3847/1538-4357/aab9bb}{\emph{Astrophys. J.}
  {\bfseries 859} (2018) 101}
  [\href{https://arxiv.org/abs/1710.00845}{{\ttfamily 1710.00845}}].

\bibitem{Pisanti:2007hk}
O.~Pisanti, A.~Cirillo, S.~Esposito, F.~Iocco, G.~Mangano, G.~Miele et~al.,
  \emph{{{PArthENoPE}}: {{Public Algorithm Evaluating}} the {{Nucleosynthesis}}
  of {{Primordial Elements}}},
  \href{https://doi.org/10.1016/j.cpc.2008.02.015}{\emph{Comput. Phys. Commun.}
  {\bfseries 178} (2008) 956}
  [\href{https://arxiv.org/abs/0705.0290}{{\ttfamily 0705.0290}}].

\bibitem{ACT:2020frw}
{\scshape ACT} collaboration, \emph{{The Atacama Cosmology Telescope: a
  measurement of the Cosmic Microwave Background power spectra at 98 and 150
  GHz}}, \href{https://doi.org/10.1088/1475-7516/2020/12/045}{\emph{J. Cosmol.
  Astropart. Phys.} {\bfseries 12} (2020) 045}
  [\href{https://arxiv.org/abs/2007.07289}{{\ttfamily 2007.07289}}].

\bibitem{ACT:2020gnv}
{\scshape ACT} collaboration, \emph{{The Atacama Cosmology Telescope: DR4 Maps
  and Cosmological Parameters}},
  \href{https://doi.org/10.1088/1475-7516/2020/12/047}{\emph{J. Cosmol.
  Astropart. Phys.} {\bfseries 12} (2020) 047}
  [\href{https://arxiv.org/abs/2007.07288}{{\ttfamily 2007.07288}}].

\bibitem{SPT-3G:2021eoc}
{\scshape SPT-3G} collaboration, \emph{{Measurements of the E-mode polarization
  and temperature-E-mode correlation of the CMB from SPT-3G 2018 data}},
  \href{https://doi.org/10.1103/PhysRevD.104.022003}{\emph{Phys. Rev. D}
  {\bfseries 104} (2021) 022003}
  [\href{https://arxiv.org/abs/2101.01684}{{\ttfamily 2101.01684}}].

\bibitem{BICEP:2021xfz}
{\scshape BICEP, Keck} collaboration, \emph{{Improved Constraints on Primordial
  Gravitational Waves using Planck, WMAP, and BICEP/Keck Observations through
  the 2018 Observing Season}},
  \href{https://doi.org/10.1103/PhysRevLett.127.151301}{\emph{Phys. Rev. Lett.}
  {\bfseries 127} (2021) 151301}
  [\href{https://arxiv.org/abs/2110.00483}{{\ttfamily 2110.00483}}].

\bibitem{Raveri:2018wln}
M.~Raveri and W.~Hu, \emph{{Concordance and Discordance in Cosmology}},
  \href{https://doi.org/10.1103/PhysRevD.99.043506}{\emph{Phys. Rev. D}
  {\bfseries 99} (2019) 043506}
  [\href{https://arxiv.org/abs/1806.04649}{{\ttfamily 1806.04649}}].

\bibitem{Planck:2013nga}
{\scshape Planck} collaboration, \emph{{Planck intermediate results. XVI.
  Profile likelihoods for cosmological parameters}},
  \href{https://doi.org/10.1051/0004-6361/201323003}{\emph{Astron. Astrophys.}
  {\bfseries 566} (2014) A54}
  [\href{https://arxiv.org/abs/1311.1657}{{\ttfamily 1311.1657}}].

\bibitem{Herold:2021ksg}
L.~Herold, E.G.M.~Ferreira and E.~Komatsu, \emph{{New Constraint on Early Dark
  Energy from Planck and BOSS Data Using the Profile Likelihood}},
  \href{https://doi.org/10.3847/2041-8213/ac63a3}{\emph{Astrophys. J. Lett.}
  {\bfseries 929} (2022) L16}
  [\href{https://arxiv.org/abs/2112.12140}{{\ttfamily 2112.12140}}].

\bibitem{Reeves:2022aoi}
A.~Reeves, L.~Herold, S.~Vagnozzi, B.D.~Sherwin and E.G.M.~Ferreira,
  \emph{{Restoring cosmological concordance with early dark energy and massive
  neutrinos?}},  \href{https://arxiv.org/abs/2207.01501}{{\ttfamily
  2207.01501}}.

\bibitem{Joudaki:2019pmv}
S.~Joudaki et~al., \emph{{KiDS+VIKING-450 and DES-Y1 combined: Cosmology with
  cosmic shear}},
  \href{https://doi.org/10.1051/0004-6361/201936154}{\emph{Astron. Astrophys.}
  {\bfseries 638} (2020) L1}
  [\href{https://arxiv.org/abs/1906.09262}{{\ttfamily 1906.09262}}].

\bibitem{DES:2022qpf}
{\scshape DES} collaboration, \emph{{Dark Energy Survey Year 3 results:
  cosmological constraints from the analysis of cosmic shear in harmonic
  space}}, \href{https://doi.org/10.1093/mnras/stac1826}{\emph{Mon. Not. Roy.
  Astron. Soc.} {\bfseries 515} (2022) 1942}
  [\href{https://arxiv.org/abs/2203.07128}{{\ttfamily 2203.07128}}].

\bibitem{Guth:1980zm}
A.H.~Guth, \emph{{The Inflationary Universe: A Possible Solution to the Horizon
  and Flatness Problems}},
  \href{https://doi.org/10.1103/PhysRevD.23.347}{\emph{Phys. Rev. D} {\bfseries
  23} (1981) 347}.

\bibitem{Linde:1981mu}
A.D.~Linde, \emph{{A New Inflationary Universe Scenario: A Possible Solution of
  the Horizon, Flatness, Homogeneity, Isotropy and Primordial Monopole
  Problems}}, \href{https://doi.org/10.1016/0370-2693(82)91219-9}{\emph{Phys.
  Lett. B} {\bfseries 108} (1982) 389}.

\bibitem{Planck:2018jri}
{\scshape Planck} collaboration, \emph{{Planck 2018 results. X. Constraints on
  inflation}}, \href{https://doi.org/10.1051/0004-6361/201833887}{\emph{Astron.
  Astrophys.} {\bfseries 641} (2020) A10}
  [\href{https://arxiv.org/abs/1807.06211}{{\ttfamily 1807.06211}}].

\bibitem{Lucchin:1984yf}
F.~Lucchin and S.~Matarrese, \emph{{Power Law Inflation}},
  \href{https://doi.org/10.1103/PhysRevD.32.1316}{\emph{Phys. Rev. D}
  {\bfseries 32} (1985) 1316}.

\bibitem{Liddle:1988tb}
A.R.~Liddle, \emph{{Power Law Inflation With Exponential Potentials}},
  \href{https://doi.org/10.1016/0370-2693(89)90776-4}{\emph{Phys. Lett. B}
  {\bfseries 220} (1989) 502}.

\bibitem{DAmico:2022agc}
G.~D'Amico and N.~Kaloper, \emph{{Power-law Inflation Satisfies Penrose's Weyl
  Curvature Hypothesis}},  \href{https://arxiv.org/abs/2208.01048}{{\ttfamily
  2208.01048}}.

\bibitem{Langlois:2004nn}
D.~Langlois and F.~Vernizzi, \emph{{Mixed inflaton and curvaton
  perturbations}},
  \href{https://doi.org/10.1103/PhysRevD.70.063522}{\emph{Phys. Rev. D}
  {\bfseries 70} (2004) 063522}
  [\href{https://arxiv.org/abs/astro-ph/0403258}{{\ttfamily
  astro-ph/0403258}}].

\bibitem{Ferrer:2004nv}
F.~Ferrer, S.~Rasanen and J.~Valiviita, \emph{{Correlated isocurvature
  perturbations from mixed inflaton-curvaton decay}},
  \href{https://doi.org/10.1088/1475-7516/2004/10/010}{\emph{J. Cosmol.
  Astropart. Phys.} {\bfseries 10} (2004) 010}
  [\href{https://arxiv.org/abs/astro-ph/0407300}{{\ttfamily
  astro-ph/0407300}}].

\bibitem{Easson:2010uw}
D.A.~Easson and B.A.~Powell, \emph{{Optimizing future experimental probes of
  inflation}}, \href{https://doi.org/10.1103/PhysRevD.83.043502}{\emph{Phys.
  Rev. D} {\bfseries 83} (2011) 043502}
  [\href{https://arxiv.org/abs/1011.0434}{{\ttfamily 1011.0434}}].

\bibitem{Kinney:2012ik}
W.H.~Kinney, A.~Moradinezhad~Dizgah, B.A.~Powell and A.~Riotto, \emph{{Inflaton
  or Curvaton? Constraints on Bimodal Primordial Spectra from Mixed
  Perturbations}},
  \href{https://doi.org/10.1103/PhysRevD.86.023527}{\emph{Phys. Rev. D}
  {\bfseries 86} (2012) 023527}
  [\href{https://arxiv.org/abs/1203.0693}{{\ttfamily 1203.0693}}].

\bibitem{Fujita:2014iaa}
T.~Fujita, M.~Kawasaki and S.~Yokoyama, \emph{{Curvaton in large field
  inflation}}, \href{https://doi.org/10.1088/1475-7516/2014/09/015}{\emph{J.
  Cosmol. Astropart. Phys.} {\bfseries 09} (2014) 015}
  [\href{https://arxiv.org/abs/1404.0951}{{\ttfamily 1404.0951}}].

\bibitem{Bartolo:2002vf}
N.~Bartolo and A.R.~Liddle, \emph{{The Simplest curvaton model}},
  \href{https://doi.org/10.1103/PhysRevD.65.121301}{\emph{Phys. Rev. D}
  {\bfseries 65} (2002) 121301}
  [\href{https://arxiv.org/abs/astro-ph/0203076}{{\ttfamily
  astro-ph/0203076}}].

\bibitem{Bartolo:2003jx}
N.~Bartolo, S.~Matarrese and A.~Riotto, \emph{{On nonGaussianity in the
  curvaton scenario}},
  \href{https://doi.org/10.1103/PhysRevD.69.043503}{\emph{Phys. Rev. D}
  {\bfseries 69} (2004) 043503}
  [\href{https://arxiv.org/abs/hep-ph/0309033}{{\ttfamily hep-ph/0309033}}].

\bibitem{Boylan-Kolchin:2022kae}
M.~Boylan-Kolchin, \emph{{Stress Testing $\Lambda$CDM with High-redshift Galaxy
  Candidates}},  \href{https://arxiv.org/abs/2208.01611}{{\ttfamily
  2208.01611}}.

\bibitem{Amon:2022ycy}
A.~Amon et~al., \emph{{Consistent lensing and clustering in a low-$S_8$
  Universe with BOSS, DES Year 3, HSC Year 1 and KiDS-1000}},
  \href{https://arxiv.org/abs/2202.07440}{{\ttfamily 2202.07440}}.

\end{thebibliography}\endgroup

\end{document}